\documentclass[aps,prd,showpacs,superscriptaddress,groupedaddress,nofootinbib]{revtex4-2}
\usepackage{graphicx}  % needed for figures
\usepackage{dcolumn}   % needed for some tables
\usepackage{bm}        % for math
\usepackage{amssymb}   % for math
\usepackage{amsmath}
\usepackage{xcolor}
\usepackage{amsthm}
\usepackage{subfig}
\usepackage{subfloat}
\usepackage{actuarialsymbol}
\usepackage{braket}
\usepackage{mathtools}
\usepackage{bbold}
\usepackage{amssymb}
\usepackage{enumitem}
\usepackage{comment}
\newcommand{\be}{\begin{equation}}\newcommand{\ee}{\end{equation}}
\newcommand{\bea}{\begin{eqnarray}}\newcommand{\eea}{\end{eqnarray}}
\newcommand{\brr}{\begin{array}}\newcommand{\err}{\end{array}}
\newcommand{\bit}{\begin{itemize}}\newcommand{\eit}{\end{itemize}}
\newcommand{\ben}{\begin{enumerate}}\newcommand{\een}{\end{enumerate}}

\newcommand{\ba}{\begin{array}}
\newcommand{\ea}{\end{array}}
\newcommand{\ide}{1\hspace{-1mm}{\rm I}}

\definecolor{darkred}{rgb}{.8,0,0}

\definecolor{darkblue}{rgb}{0,0,.7}

\def\lf{\left}

\def\non{\nonumber}

\def\ri{\right}

\def\1{{_{1}}}\def\2{{_{2}}}

\def\noHe0{:\;\!\!\;\!\!:H_e(0):\;\!\!\;\!\!:}
\def\noHm0{:\;\!\!\;\!\!:H_\mu(0):\;\!\!\;\!\!:}

\def\lf{\left}

\def\non{\nonumber}

\def\ri{\right}

\def\1{{_{1}}}\def\2{{_{2}}}

\begin{document}
%%%%%%%%%%%%%%%%%%%%%%%%%%%%%%%%%%%%%%%%%%%%%%%%%%%%%%%%%%%%%%%%%%%%%%%%%%%%%%%%%%%%%%%%%%%%%
\title{Entanglement distribution in Bhabha scattering with entangled spectator particle}
%%%%%%%%%%%%%%%%%%%%%%%%%%%%%%%%%%%%%%%%%%%%%%%%%%%%%%%%%%%%%%%%%%%%%%%%%%%%%%%%%%%%%%%%%%%%

\author{Massimo Blasone, Gaetano Lambiase and Bruno Micciola}
\email{blasone@sa.infn.it}
\affiliation{Dipartimento di Fisica, Universit\`a di Salerno, Via Giovanni Paolo II, 132 I-84084 Fisciano (SA), Italy}
\email{lambiase@sa.infn.it}
\affiliation{INFN, Sezione di Napoli, Gruppo collegato di Salerno, Italy}
\email{bmicciola@unisa.it}

\date{\today}

\begin{abstract}
%%%%%%%%%%%%%%%%%%%%%%%%%%%%%%%%%%

We analyze how entanglement is generated and distributed in a Bhabha scattering process $(e^-e^+\rightarrow e^-e^+)$ at tree level. In our setup an electron $A$ scatters with a positron $B$,  which is initially entangled with another electron $C$ (spectator), that does not participate directly to the process. 
%The analysis proceeds by studying correlations between helicity states, in terms of concurrence, in each bipartite subsystems $AB$, $AC$ and $BC$, called channels. After studying the density matrix structure of $C$, 
We find that the QED scattering generates and distributes entanglement in a non-trivial way among the three particles: the  correlations in the output channels  $AB$, $AC$ and $BC$ are studied in detail as functions of the scattering parameters and of the initial entanglement weight. 
%the  depending on the configuration of the parameters $\theta$, the scattering angle, $\mu$, the ratio between the particles incoming momentum in the center of mass reference frame, and $\eta$, the entanglement weight. 
%Among all the results, we find that for some configuration of $\theta$ and $\eta$, in the relativistic regime $A$ and $C$, that do not interact, result perfectly correlated. For others the QED interaction generates  entanglement in $BC$ channel, where it was already present, and in the other two. For still others, entanglement is transfers from $BC$ to $AC$ and $AB$ channels. 
%These generation and distributions of correlations, that arise and can be regulated by a fundamental interaction, represent a first step towards a deeper comprehension of the nature of the entanglement and applications in the context of Quantum Information and QFT theories. 
Although derived in a specific case, our results exhibit some general features of other similar QED scattering processes, for which the extension of the present analysis is straightforward.
%%%%%%%%%%%%%%%%%%%%%%%%%%%%%%%%%
\end{abstract}

\vskip -1.0 truecm

\maketitle

%%%%%%%%%%%%%%%%%%%%%%%%
\section{Introduction}
%%%%%%%%%%%%%%%%%%%%%%
The r\^ole of entanglement in high-energy physics has recently become a very active research area \cite{Bertlmann:2001sk}-\cite{Sinha:2022crx}. Much attention has been devoted to the study of  quantum correlations  in neutrino oscillations \cite{Blasone:2007wp}-\cite{Blasone:2022ete}, since neutrinos are regarded as possible alternative carriers of quantum information with respect to photons. Various aspects of entanglement in scattering processes have been investigated in the context of different fundamental interactions \cite{Beane:2018oxh}-\cite{Fonseca:2021uhd}. In Refs \cite{Afik:2020onf}-\cite{Afik:2022dgh} entanglement and other types of quantum correlations have been studied in top-antitop quark pairs, using the experimental data from proton-proton and proton-antiproton collision at the LHC.

In Refs.\cite{Cervera-Lierta:2017tdt, CerveraLierta:2019ejf} it is analyzed how maximal entanglement is generated at the fundamental level in QED by studying correlations between helicity states at tree-level for various scattering processes. In particular, the authors describe the mechanisms that generate maximal entanglement and its relation with the scattering amplitudes in the high energy regime.
%This condition is then promote to a fundamental principle in the context of unconstrained QED and it is investigated how this constrains the QED and the weak interaction couplings. In the case of photon-electron interactions unconstrained by gauge symmetry, they find that this requirement allows to reproduce QED. For Z-mediated weak scattering, the maximal entanglement principle leads to non-trivial predictions for the value of the weak mixing angle $\theta_W$. 
An extension of this work is given in Ref.\cite{Fedida:2022izl}, where the entanglement generation was analyzed at all energies for pure and mixed final states for arbitrary initial mixtures of helicity states. It was shown that maximal entanglement can originate in all situations where one dominating channel leads to balanced superpositions of helicity states. It was also found that loop corrections do not significantly alter the results obtained at tree level. 

An interesting extension of these studies is given in Refs.\cite{Araujo:2019mni,Fonseca:2021uhd}. In Ref.\cite{Araujo:2019mni} it is considered a QED scattering of two particles $A$ and $B$, in which $B$ is initially entangled with a third particle $C$ that does not participate directly in the process. The authors investigate the effects of the scattering both on the particle $C$ and in the  bipartite channels. In Ref.\cite{Fonseca:2021uhd} the model is extended by considering a general three-partite entangled state in input and applied to the case of QED inelastic tree-level process $e^-e^+\rightarrow \mu^-\mu^+$. Further extensions of these works are given in  Refs.\cite{Fan:2021qfd,Shivashankara:2023koj}.

\begin{figure}
    \centering
    \includegraphics[width =5.6 cm]{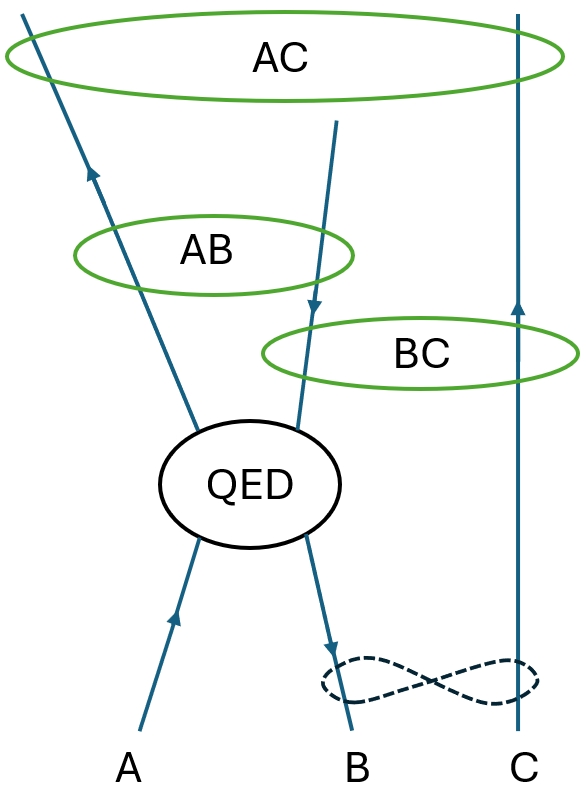}
    \caption{Schematic representation of the process considered in this work. Particles $A$ and $B$ scatter through a QED process at tree-level. Particle $C$ (spectator) does not participate to the scattering and is initially entangled to particle $B$. The ellipses $AB$, $AC$ and $BC$ represent the bi-partitions for which we calculate the entanglement.}
    \label{fig scheme}
\end{figure}

In this work, starting from the same framework used in Ref.\cite{Araujo:2019mni}, we study in detail the case of Bhabha scattering, describing how entanglement is generated and distributed in the three bipartite sub-systems $AB$, $AC$ and $BC$ after the scattering as represented in Fig.\ref{fig scheme}. This process depends on the  value of the initial entanglement weight $\eta$  and of the scattering parameters $\theta$ (scattering angle) and $\mu$ (the ratio between $\vec{p}$ the incoming momentum of $e^-$ and $e^+$ in the CM reference frame and m the particles mass). For incoming momenta of the order of the mass, the entanglement has a non trivial distribution in the three output channels.
On the other hand, in the relativistic regime, for specific values of parameters, entanglement transfers completely from the $BC$ bipartition (where initially was present) to the $AC$  bipartition, as a consequence of the scattering between $A$ and $B$.

The paper is organized as follows: in  Section II we set up the problem and analyze the density matrix structure for the  $C$ subsystem, relative to the spectator particle, before and after the scattering. In Section III we study the entanglement generation in the scattering between $A$ and $B$, taking $B$ as a superposition of helicity states. The study of such a reference state is useful for the analysis carried out in Section IV, where we consider  the entanglement generation and distribution  in the presence of a spectator particle $C$. Section V is devoted to conclusions and outlook.

%%%%%%%%%%%%%%%%%%%%%%
\section{Entanglement in QED scattering with spectator particles}
%%%%%%%%%%%%%%%%%%%%%%

In this Section we briefly recall the setup used in Ref.\cite{Araujo:2019mni}, in which it is considered a general QED scattering process at tree level involving two particles ($AB \rightarrow AB$), where $B$, before the interaction, is entangled in spin with a third particle $C$ that does not participate to the process, as schematically shown in Fig.\ref{fig scheme}. For simplicity, we perform our calculations in the COM  reference frame for particles $A$ and $B$ and assume that the spectator momentum ${\bf q}$ is aligned in the same direction of the incoming momenta of $A$ and $B$.

The internal product of fermion states is defined as
\begin{equation}
\label{intprod}
    \braket{k,a|p,b} = 2E_{\bf k}(2\pi)^3 \delta^{(3)}({\bf k}-{\bf p})\delta_{a,b},
\end{equation}
where $k$ and $p$ are the 4-momenta and $a$ and $b$ are the spin indices.

The initial state  is taken to be:
\begin{equation}
\label{instate}
    \ket{i} = \ket{p_{1},a}_A \otimes \Big(\cos\eta \ket{p_{2}, \uparrow}_B\otimes \ket{q, \uparrow}_C + e^{i\beta}\sin\eta \ket{p_{2}, \downarrow}_B \otimes \ket{q, \downarrow}_C\Big),
\end{equation}
whose final state, is given by
\begin{eqnarray}
\label{finstate} \nonumber
    \ket{f} &=& \ket{i} + i\sum_{r,s}\int \frac{d^3 {\bf p_3} d^3{\bf  p_4}}{(2\pi)^6 2E_{\bf p_3}2E_{\bf p_4}}\delta^{(4)}(p_1+p_2-p_3-p_4)\Big[\cos \eta \mathcal{M}(a,\uparrow;r,s)\ket{p_3, r}_A \otimes \ket{p_4, r}_B \otimes \ket{q, \uparrow}_C \\
      && \qquad \qquad \qquad +\,  e^{i\beta}\sin \eta \mathcal{M}(a,\downarrow;r,s)\ket{p_3, r}_A \otimes \ket{p_4, r}_B \otimes \ket{q, \downarrow}_C\Big].
\end{eqnarray}

The partial trace operation is given by
\begin{equation}
\label{trace}
    Tr_X[{\rho}] = \sum_{\sigma}\int \frac{d^3\bf k}{(2\pi)^32E_{\bf k}} \bigr(\ide_r \otimes \prescript{}{X}{\bra{k,\sigma}}\bigl)\rho\bigr(\ide_r \otimes \ket{k, \sigma}_X\bigl),
\end{equation}
where $\ide_r$ denotes the identity operation in the remaining subspaces, $k$ and $\sigma$ are the 4-momentum and spin indices as before and $X$ is the generic space with respect to which we calculate the trace.

In Eq.\eqref{finstate}, $\mathcal{M}(a,\uparrow;r,s)$ represents the scattering amplitude $\mathcal{M}(p_1,a,p_2,\uparrow\,;\, p_3,r,p_4,s)$ and the same for $\mathcal{M}(a,\downarrow;r,s)$, where we have omitted initial and final momenta for shortness.
So, we can describe the final states in terms of the density matrix:
\begin{equation}
\label{rhofin}
    \rho_{ABC}^{f} = \frac{1}{\mathcal{N}}\, \ket{f}\bra{f}.
\end{equation}
where $\mathcal{N}$ is the normalization constant.
Using Eq.\eqref{trace} for evaluating Eq.\eqref{rhofin} and applying the following relations
\begin{equation}
\label{Tfactor}
    2\pi\delta^{(0)}(E_i-E_f) = \int_{-T/2}^{T/2} e^{i(E_i-E_f)} dt,
\end{equation}

\begin{equation}
\label{Vfactor}
        (2\pi)^3\delta^{(3)}({\bf k}-{\bf p}) =  V\delta_{\bf{k},\bf{p}} \hspace{0.1cm},
\end{equation}
which imply that $(2\pi)\delta^{(0)}(0)=T$ and $(2\pi)^3\delta^{(3)}(0)=V$, we can easily compute the normalization constant $\mathcal{N}$:
\begin{equation}
\label{norm}
    \mathcal{N} = Tr_A[Tr_B[Tr_C(\rho_{ABC}^{f})]] = 2E_{p_1}2E_{p_2}V^2 + T^2V\Lambda ,
\end{equation}
where
\begin{equation}
\label{lambda}
    \Lambda = \int \frac{d^3\bf p_3}{(2\pi)^{12} 2E_{\bf p_3} 2E_{\bf p_3-\bf p_1-\bf p_2}} \sum_{r,s}\Big(\cos^2 \eta\, |\mathcal{M}(a,\uparrow;r,s)|^2 + \sin^2 \eta \, |\mathcal{M}(a,\downarrow;r,s)|^2\Big)\big|_{\bf p_4=\bf p_3-\bf p_1-\bf p_2}.
\end{equation}

\subsection{Effects of scattering on spectator particle}

%An important outcome of the analysis of Ref.\cite{Araujo:2019mni} is the investigation of  
We now investigate the effects of the scattering over the spectator particle $C$.  To this aim, we consider the reduced density matrix
\begin{equation}\label{rhoGae}
\rho_C\equiv \frac{1}{\mathcal{N}}Tr_A[Tr_B[\rho_{ABC}]]
\end{equation}
relative to the $C$ sub-system, both for the initial and the final states.

By using Eq.\eqref{trace} to calculate the reduced density matrix for $C$ from the initial state  Eq.\eqref{instate}, we get
\begin{equation}
\label{rhocin}
   \rho_{C}^{i} = 2E_{p_1}2E_{p_2}\Big(\cos^2{\eta}\ket{\uparrow}_C \prescript{}{C}{\bra{\uparrow}}+\sin^2{\eta}\ket{\downarrow}_C \prescript{}{C}{\bra{\downarrow}}\Big) \otimes\ket{q}_C \prescript{}{C}{\bra{q}}.
\end{equation}
From now on, as the entanglement is consider over the spin degrees of freedom, we omit the factorized part $\ket{q}_C\prescript{}{C}{\bra{q}}$ of the spectator momentum subspace. 
The initial $C$-density matrix can be express in matrix form as
\be
\rho_{C}^{i} = \begin{pmatrix}
   \cos^2{\eta} &  0 \\
   0 & \sin^2{\eta}
\end{pmatrix} .
\ee

\medskip

The reduced density matrix of the final state for $C$ reads:
\begin{eqnarray}
\label{rhocfinus} \non
     \rho_C^f &=& \frac{1}{\mathcal{N}}\sum_{\sigma\sigma'}\int\frac{d^3k d^3k'}{(2\pi)^6 2E_k 2E_{k'}}\Big(\mathbb{1}_r\otimes \prescript{}{B}{\bra{k'\sigma'}}\prescript{}{A}{\bra{k\sigma}})\ket{f}\bra{f}(\ket{k\sigma}_A\ket{k'\sigma'}_B\otimes\mathbb{1}_r\Big) 
     \\  \non \vspace{0.1cm}
     &=&\frac{1}{\mathcal{N}}\biggl\{2E_{p_1}2E_{p_2}V^2\Big(\cos^2{\eta}\ket{\uparrow}_C \prescript{}{C}{\bra{\uparrow}}+\sin^2{\eta}\ket{\downarrow}_C\prescript{}{C}{\bra{\downarrow}}\Big)
     \\  \non
    &+&\!\!T^2V \!\!\!\int\!\!\frac{d^3p_3}{(2\pi)^{3} 2E_{p_3} 2E_{p_1+p_2-p_3}} \biggl[\sum_{rs}\Big(\cos^2{\eta}|\mathcal{M}(a,\uparrow;r,s)|^2\ket{\uparrow}_C\prescript{}{C}{\bra{\uparrow}} \vspace{0.1cm}
     +e^{-i\beta}\cos{\eta}\sin{\eta}\mathcal{M}(a,\uparrow;r,s)\mathcal{M^{\dagger}}(a,\downarrow;r,s)\ket{\uparrow}_C\prescript{}{C}{\bra{\downarrow}}
     \\
    &+&e^{i\beta}\cos{\eta}\sin{\eta}\mathcal{M}(a,\downarrow;r,s)\mathcal{M^{\dagger}}(a,\uparrow;r,s)\ket{\downarrow}_C\prescript{}{C}{\bra{\uparrow}} \vspace{0.2cm}
     + \sin^2 {\eta} |\mathcal{M}(a,\downarrow;r,s)|^2\ket{\downarrow}_C\prescript{}{C}{\bra{\downarrow}}\Big)\biggr]\biggr\}\otimes \ket{q}_C\prescript{}{C}{\bra{q}}.
\end{eqnarray}

This, in matrix form as before, looks like
\begin{equation}\label{rhocfinmatus}
\hspace{-2mm}\rho_{C}^f=\frac{1}{\mathcal{N}} 
\lf(\ba{cc}
\bigl(2E_{p_1}2E_{p_2}V^2+\int_{\bf{p}_3}\sum_{rs}|\mathcal{M}(a,\uparrow;r,s)|^2\bigr)\cos^2{\eta} & \hspace{0.2cm} e^{-i\beta}\cos{\eta}\sin{\eta}\int_{\bf{p}_3}\sum_{rs} \mathcal{M}(a,\uparrow;r,s)\mathcal{M^{\dagger}}(a,\downarrow;r,s) 
    \\ [2mm]
   e^{i\beta}\cos{\eta}\sin{\eta}\int_{\bf{p}_3}\sum_{rs} \mathcal{M}(a,\downarrow;r,s)\mathcal{M^{\dagger}}(a,\uparrow;r,s) & \bigl(2E_{p_1}2E_{p_2}V^2+\int_{\bf{p}_3}\sum_{rs}|\mathcal{M}(a,\downarrow;r,s)|^2\bigr)\sin^2{\eta}
   \ea\ri)
\end{equation}
where  we have defined the shorthand notation:
$
\int_{\bf{p}_3} \equiv T^2V\int\frac{d^3p_3}{(2\pi)^{3} 2E_{p_3} 2E_{p_1+p_2-p_3}}.
$

\smallskip

Now, fixing the incoming momentum, we can study the system in terms of the helicity states. Using the spinors reported in Appendix A and considering  the specific case of Bhabha scattering, we calculate the scattering amplitude and the elements of the matrix $\rho_C^{f}$,  with an arbitrary  initial polarized state for $e^+$ and $e^-$.

The off-diagonal terms in Eq.\eqref{rhocfinmatus}
vanish identically\footnote{In Ref.\cite{Shivashankara:2023koj}, in the context of Compton scattering, by resorting to unitarity and to the optical theorem, a similar conclusion is obtained.}: the expression $\sum_{rs}\mathcal{M}(a,\uparrow;r,s)\mathcal{M^{\dagger}}(a,\downarrow;r,s)$ is an odd function of the scattering angle $\theta$. 

The diagonal terms, because of the normalization constant $\mathcal{N}$, are the same of $\rho_C^{i}$. Thus we have
\begin{equation}
\label{rhoineqrhofin}
    \rho_C^{i} = \rho_C^{f} = \begin{pmatrix}
       \cos^2{\eta} &  0 \\
    0 & \sin^2{\eta} 
    \end{pmatrix}.
\end{equation}

The same conclusion holds for the unpolarized case and for other QED scattering processes. This ensures that if $C$ is out of the light cone of the scattering event, no superluminal communication can occur: as a consequence,   the $C$-observer cannot get any information about the scattering process, whatever the observable he/she measures. This result is in contrast with the one reported in Ref.\cite{Araujo:2019mni} in which it is found that, as a consequence of the scattering between $A$ and $B$, off-diagonal terms of $\rho^{f}_C$ appear, with the implication of nonzero expectation values for some observables, like $\braket{\sigma_x}$ or $\braket{\sigma_y}$.

\vspace{0.5cm}

\section{Entanglement measure and reference system}

 In Refs.\cite{Cervera-Lierta:2017tdt,CerveraLierta:2019ejf}  QED scattering processes at tree-level  are considered with disentangled incoming particles $A$ and $B$ in both cases with $RR$ and $RL$ initial helicity states. Then, after the scattering, the generated entanglement is quantified by using concurrence. In our setup, represented in Fig.\ref{fig scheme}, the scattering involves an electron $A$ that collides with an entangled state formed by a positron $B$ and an electron $C$. The scattering amplitudes and the spinors are described as functions of the scattering angle $\theta$ (see Appendices A and B). 
 
 In order to better understand the generation and distribution of the entanglement, mediated by the QED interaction, we preliminary analyze the case of the (Bhabha) scattering of an electron $A$ and a positron $B$ in which the latter is in a superposition of helicity states (see Eq.\eqref{inrefstate} below). This system  will be used as a benchmark in Section IV to compare the generation and transfer of the entanglement in the various channels when the spectator particle $C$ is added. 
 
\smallskip
 
 To calculate the concurrence, we use the definition as in Ref.\cite{Wootters:1997id}:
\begin{equation}
\label{conc}
    C(\rho) = max(0,\lambda_1-\lambda_2-\lambda_3-\lambda_4),
\end{equation}
where the $\lambda_i$, in decreasing order, are the square root of the eigenvalues of the matrix
\begin{equation}
\label{matconc}
    \mathcal{R}= \rho_{s_1s_2}\tilde{\rho}_{s_1s_2},
\end{equation}
with 
$
\tilde{\rho}_{s_1s_2}=(\sigma_y\otimes\sigma_y)\rho^{\ast}_{s_1s_2}(\sigma_y\otimes\sigma_y), 
$
where $\sigma_y$ is a Pauli matrix and $s_1,s_2$ are indices running in the bipartition subspaces.

\medskip

Following the above discussion, we define our \emph{initial reference state} as
%From Eq.\ref{instate}, fixing the incoming momentum, we can write up to a constant the initial state in terms of helicity states as: 
%
\begin{equation}
\label{inrefstate}
    \ket{i}_{REF} = \ket{R}_A\otimes\Big(\cos{\eta}\ket{R}_B+e^{i\beta}\sin{\eta}\ket{L}_B\Big).
\end{equation}

After the scattering, if we limit our attention to a selection of results at a fixed angle $\theta\neq 0,2\pi$, we can express, up to a normalization factor\footnote{The normalization can be fixed after the operation of momentum filtering  (i.e. selection of the measurements relative to a specific scattering angle $\theta$) that is formally described by applying a POVM operator as discussed in Ref.\cite{Fedida:2022izl}.}, the\emph{ final reference state} as
\begin{equation}
\label{finrefstate}
   \ket{f}_{REF}=\sum_{r,s=R,L}\biggl[\cos\eta\, \mathcal{M}(RR;rs)\ket{r}_A\ket{s}_B + e^{i\beta}\sin{\eta}\, \mathcal{M}(RL;rs)\ket{r}_A\ket{s}_B\biggr],
\end{equation}
where the $\mathcal{M}(\prescript{RR}{LR}{};rs)$ are the scattering amplitudes  as reported in Appendix B. These correspond to the amplitudes given in Refs.\cite{Cervera-Lierta:2017tdt,CerveraLierta:2019ejf}.

\smallskip

By using Eqs.\eqref{inrefstate},\eqref{finrefstate}, we obtain  the density matrices of initial and final reference state (omitting  normalization factors):
 \begin{eqnarray}
\label{rhorefinstate}
 \nonumber
    \hspace{-2cm}\rho^{i}_{REF} &= \Big[\cos^2{\eta}\,\ket{R}_A\ket{R}_B\prescript{}{B}{\bra{R}\prescript{}{A}{\bra{R}}}
     \nonumber
    +e^{-i\beta}\sin{\eta}\cos{\eta}\,\ket{R}_A\ket{R}_B\prescript{}{B}{\bra{L}\prescript{}{A}{\bra{R}}}
    \\ &+e^{i\beta}\sin{\eta}\cos{\eta}\,\ket{R}_A\ket{L}_B\prescript{}{B}{\bra{R}\prescript{}{A}{\bra{R}}}
    +\sin^2{\eta}\,\ket{R}_A\ket{L}_B\prescript{}{B}{\bra{L}\prescript{}{A}{\bra{R}}} \Big],
\end{eqnarray}
\begin{eqnarray}
\label{rhorefstate}
 \nonumber
    \hspace{-2cm}\rho^{f}_{REF} &= \!\!\!\sum\limits_{r,s,r^{{}_{'}},s^{{}_{'}}}&\!\!\!\!\Big[\cos^2{\eta}\,\mathcal{M}(RR,rs)\mathcal{M}^{\dagger}(RR,r^{{}_{'}}s^{{}_{'}})\ket{r}_A\ket{s}_B\prescript{}{B}{\bra{s^{{}_{'}}}}\prescript{}{A}{\bra{r^{{}_{'}}}}
    \\ \nonumber
    &&+e^{-i\beta}\sin{\eta}\cos{\eta}\,\mathcal{M}(RR,rs)\mathcal{M}^{\dagger}(RL,r^{{}_{'}}s^{{}_{'}})\ket{r}_A\ket{s}_B\prescript{}{B}{\bra{s^{{}_{'}}}}\prescript{}{A}{\bra{r^{{}_{'}}}}
    \\ \nonumber
    &&+e^{i\beta}\sin{\eta}\cos{\eta}\,\mathcal{M}(RL,rs)\mathcal{M}^{\dagger}(RR,r^{{}_{'}}s^{{}_{'}})\ket{r}_A\ket{s}_B\prescript{}{B}{\bra{s^{{}_{'}}}}\prescript{}{A}{\bra{r^{{}_{'}}}}
    \\ 
    &&+\sin^2{\eta}\,\mathcal{M}(RL,rs)\mathcal{M}^{\dagger}(RL,r^{{}_{'}}s^{{}_{'}})\ket{r}_A\ket{s}_B\prescript{}{B}{\bra{s^{{}_{'}}}}\prescript{}{A}\!{\bra{r^{{}_{'}}}} \Big].
\end{eqnarray}

By using Eqs.\eqref{conc},\eqref{matconc} we calculate the amount of entanglement in the states $\rho^{i}_{REF}$ and $\rho^{f}_{REF}$. For simplicity, we take the phase $\beta=0$. The concurrence of the initial reference state vanishes, as expected. On the other hand, the concurrence of the final state is a cumbersome expression and it is not reported here. However, in the relativistic limit, we obtain a simple form:
\bea\label{ConcRefRel}
    \lim_{\mu\rightarrow\infty}C(\rho^f_{REF})
    %&=&\frac{64\bigl(1-\cos(2\eta)\bigr)\sin^4(\theta/2)\cos^4(\theta/2)}{64\cos^2(\eta)+(35+28\cos(2\theta)+\cos(4\theta))\sin^2(\eta)}, \\
    &=&\frac{2 \sin^2\eta \sin^4(\theta/2)\cos^4(\theta/2)}{1-(1-\frac{1}{8}\sin^2\theta)\sin^2\theta\sin^2\eta}.
\eea

\begin{figure}[h]
 \subfloat[][\emph{}]{\includegraphics[width =5.55 cm]{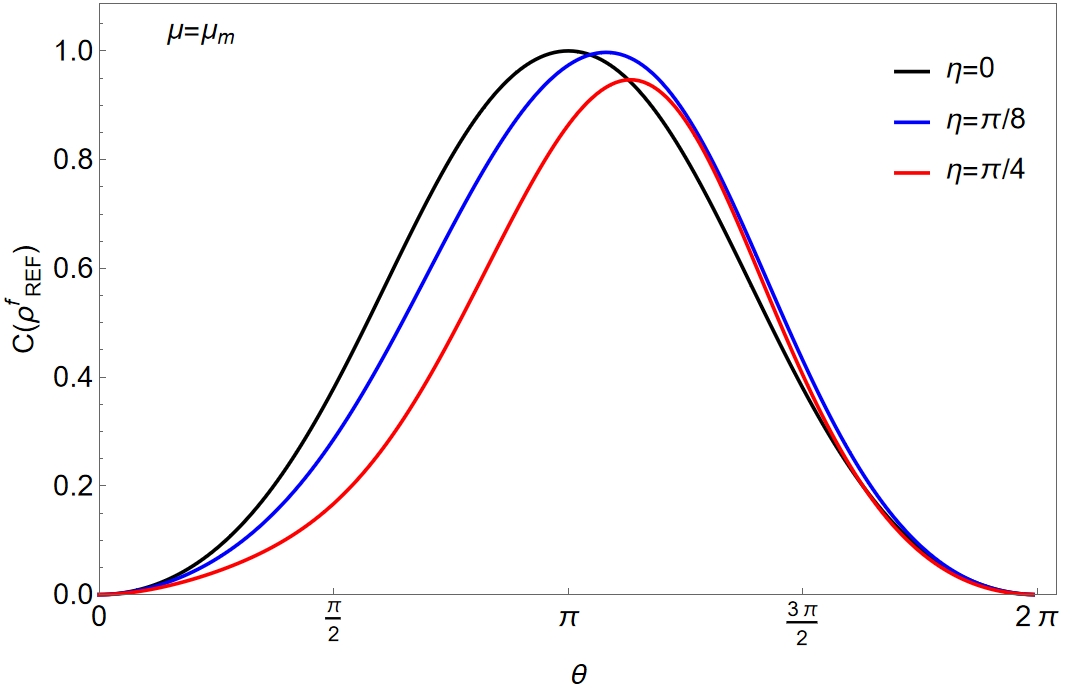}
\label{fig2a}}\quad
 \subfloat[][\emph{}]{\includegraphics[width =5.55 cm]{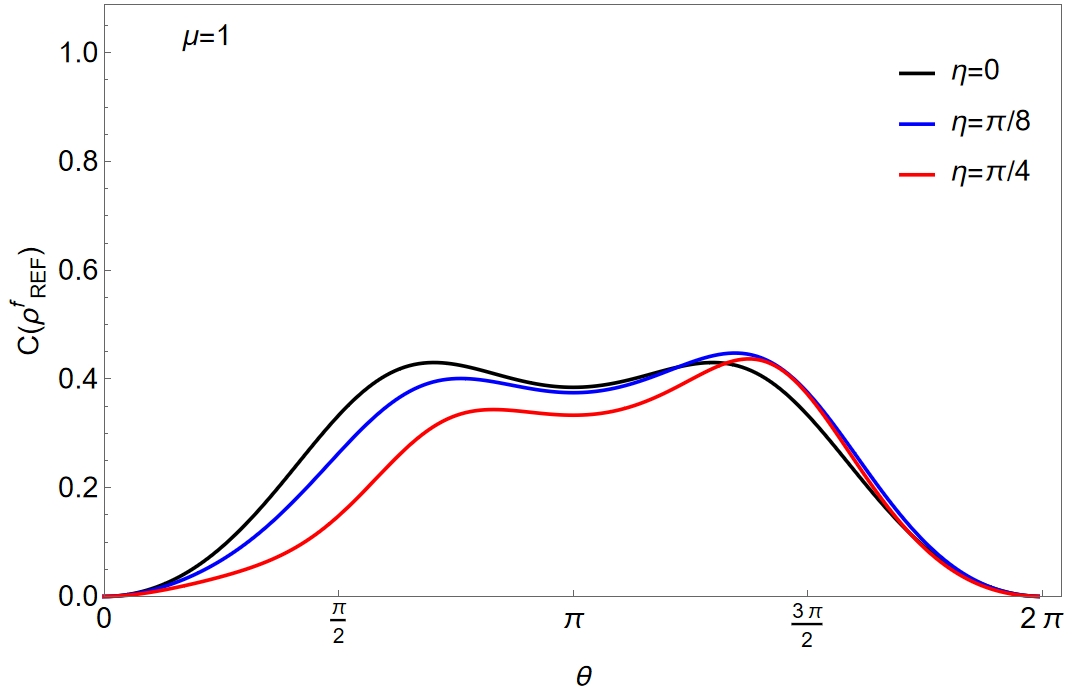}
\label{fig2b}}\quad
 \subfloat[][\emph{}]{\includegraphics[width =5.55 cm]{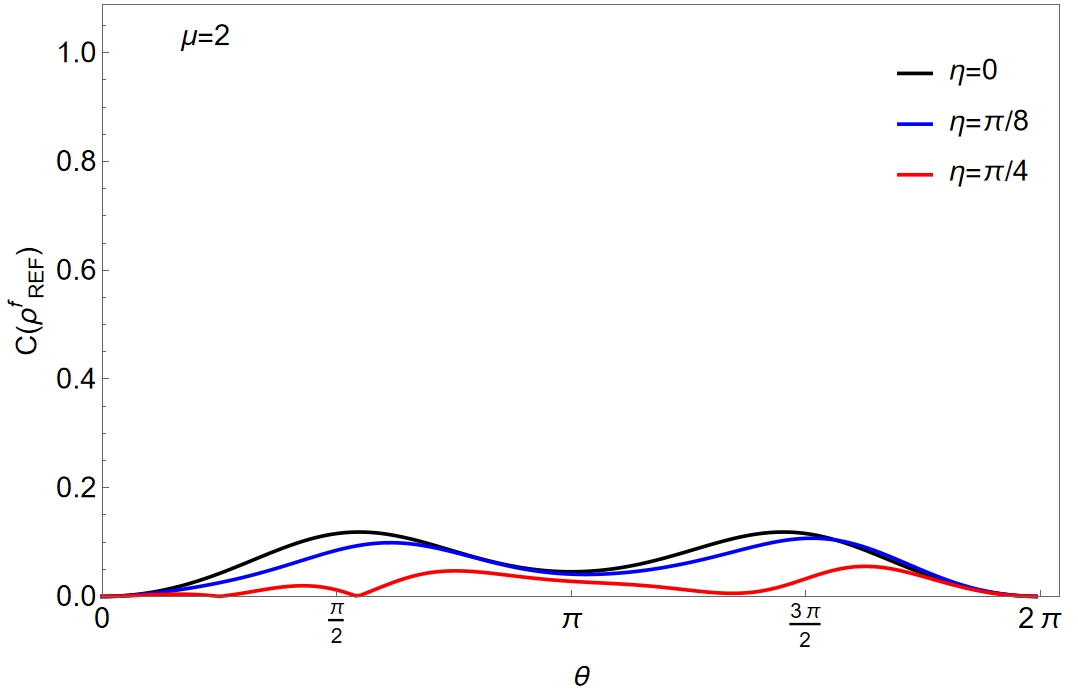}
\label{fig2c}}
\caption{Concurrence of reference state Eq.\eqref{rhorefstate} for low momenta. Black lines correspond to the case of Ref.\cite{Cervera-Lierta:2017tdt}. The value of $\mu$ in (a) maximizes the concurrence that for $\eta=0$ as in \cite{Cervera-Lierta:2017tdt}. 
%By increasing entanglement weight $\eta$, concurrence asymmetry develops. For increasing momenta, concurrence decreases.
}
\label{figtwo}
\end{figure}

\begin{figure}[h]
 \subfloat[][\emph{}]{\includegraphics[width =5.55 cm]{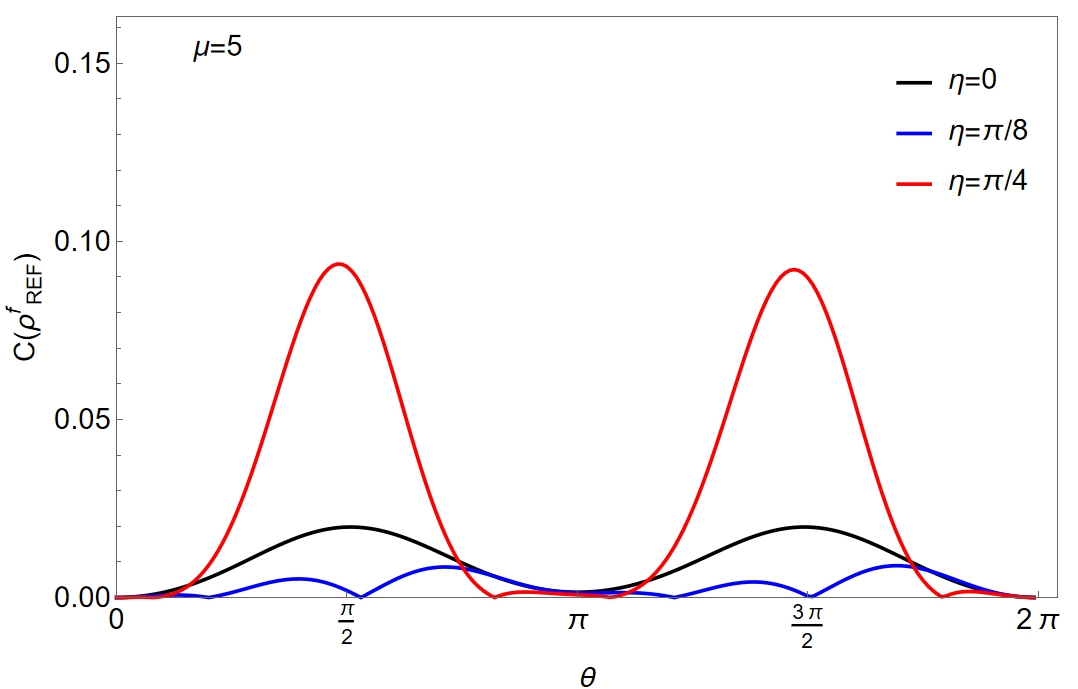}
\label{fig3a}}\quad
 \subfloat[][\emph{}]{\includegraphics[width =5.55 cm]{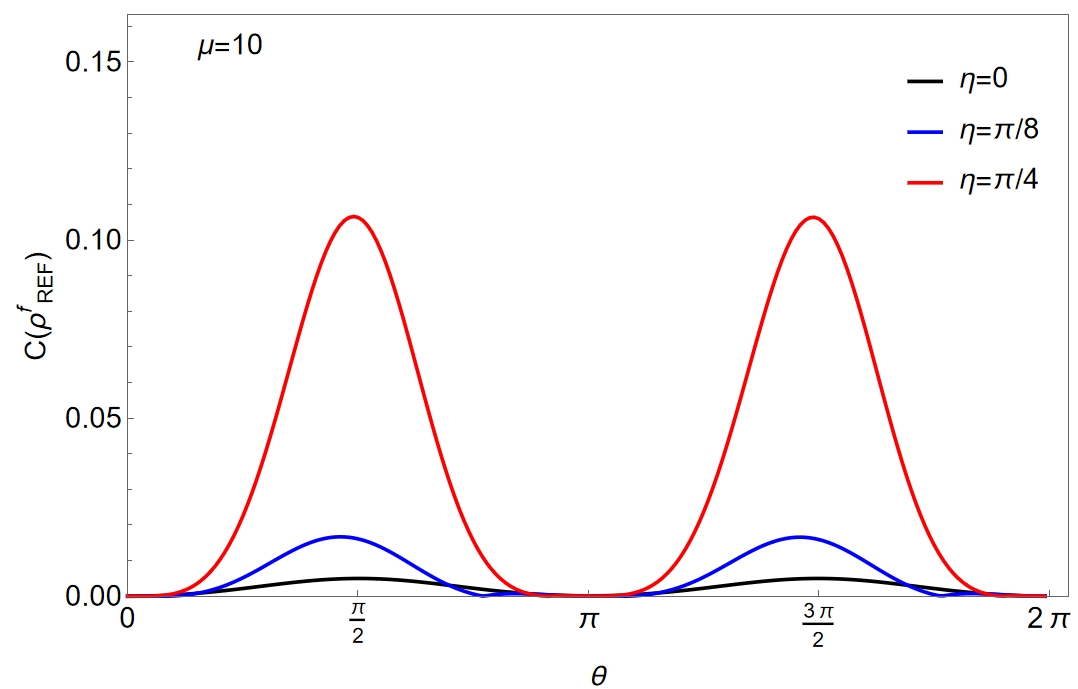}
\label{fig3b}}\quad
 \subfloat[][\emph{}]{\includegraphics[width =5.55 cm]{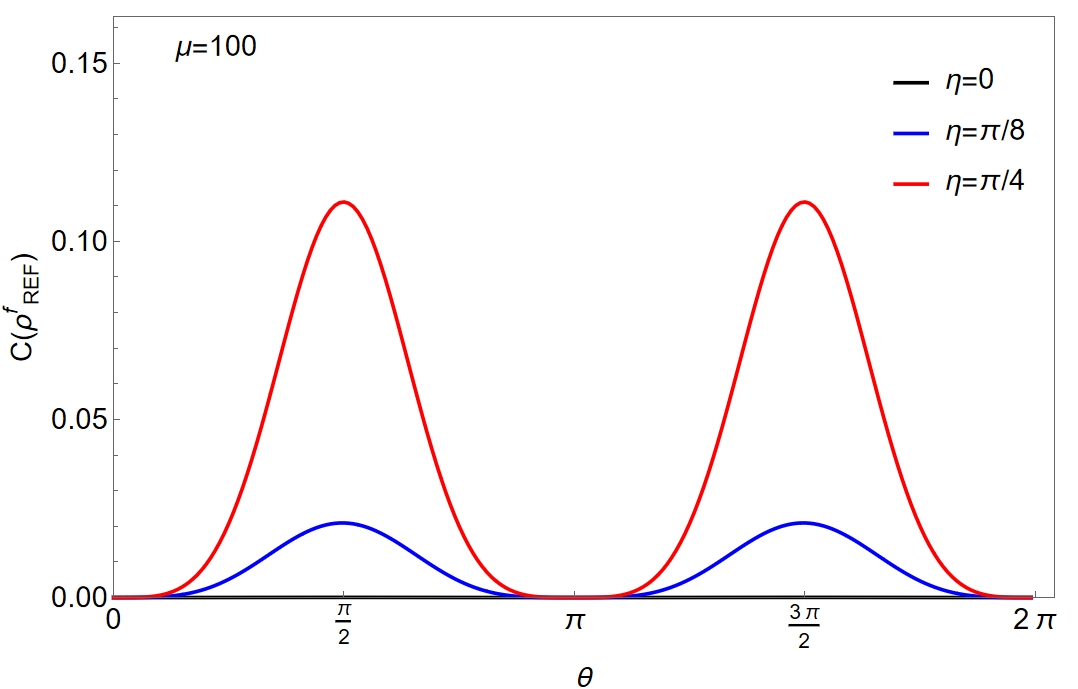}
\label{fig3c}}
\caption{Concurrence of reference state Eq.\eqref{rhorefstate} for high momenta. Black lines correspond to the case of Ref.\cite{Cervera-Lierta:2017tdt}. 
%For increasing momenta $C(\rho^{f}_{REF})$ decreases, while for entanglement weights $\eta\neq 0$ the concurrences increase forming their peaks at $\theta=\pi/2$ and $\theta=3\pi/2$ and asymmetry vanish by increasing momenta.
} 
\label{figthree}
\end{figure}

\begin{figure}[h]
 \includegraphics[width =5.55 cm]{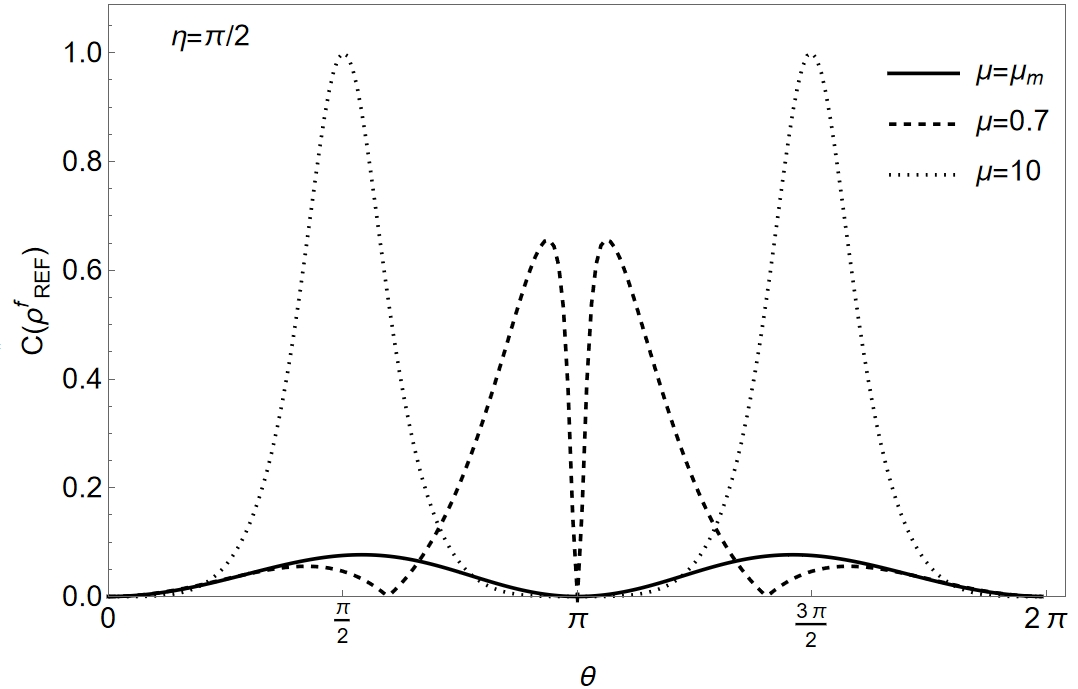}
\caption{Concurrence of reference state
Eq.\eqref{rhorefstate} for different incoming momenta and particles with $RL$ helicities ($\eta=\pi/2$). In the high energy limit maximal entanglement is generated for $\theta=\pi/2$ and $\theta=3\pi/2$ (see also Ref.\cite{Cervera-Lierta:2017tdt}).} 
\label{figfour}
\end{figure}

\begin{figure}[h]
 \subfloat[][\emph{}]{\includegraphics[width =5.0 cm]{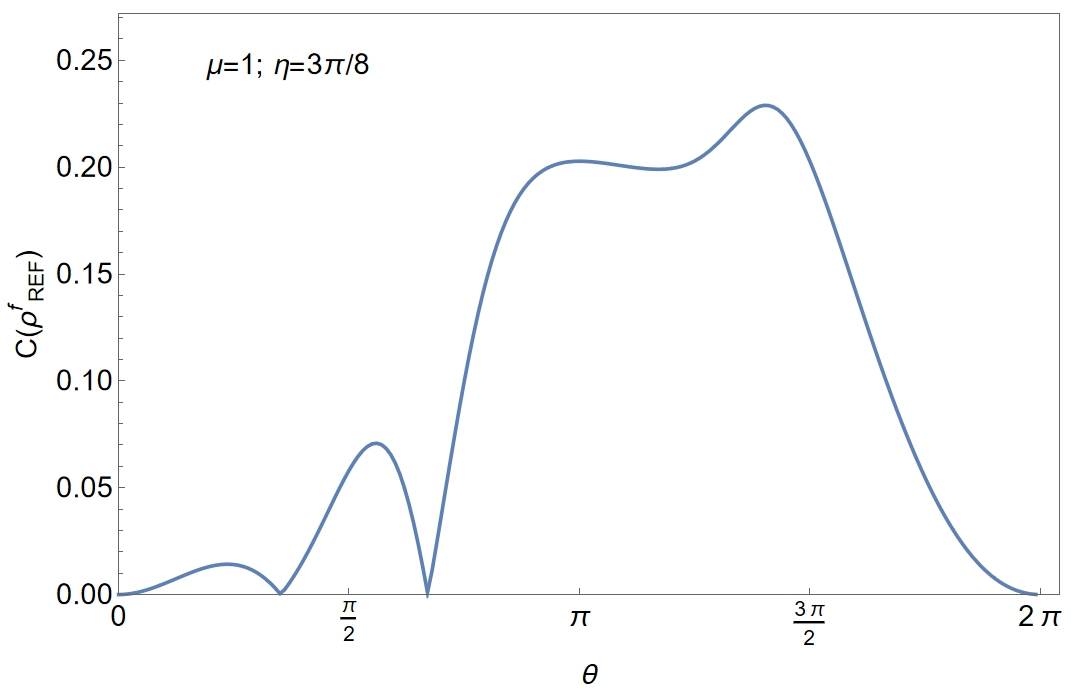}}
\label{fig5a}\quad
 \subfloat[][\emph{}]{\includegraphics[width =5.0 cm]{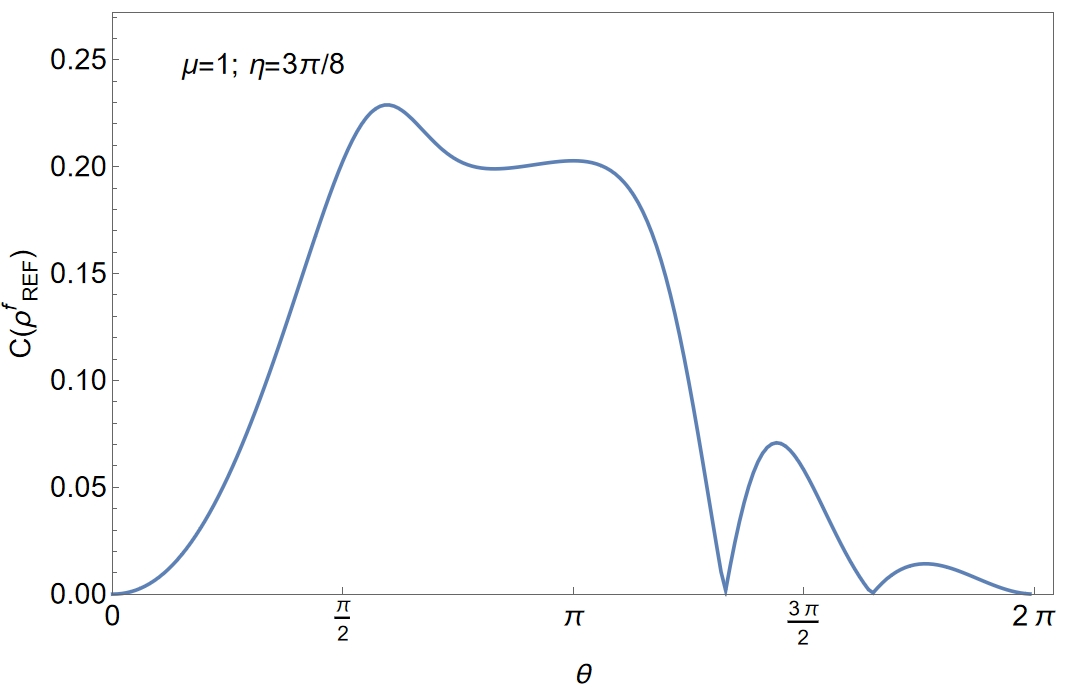}}
\label{fig5b}\quad
\caption{In (a) concurrence of reference state Eq.\eqref{rhorefstate}. In (b) concurrence of reference state Eq.\eqref{rhorefstate} with opposite helicity eigenstates for incoming particles.} 
\label{figfive}
\end{figure}

The fact that concurrence before the scattering is zero, means that the entanglement after the scattering is completely generated in the process. Some representative plots  are reported in Figs.\ref{figtwo}-\ref{figfive}.

For the plots in Fig.\ref{figtwo} and Fig.\ref{figthree} we have chosen the following parameters: $\eta=\{0,\pi/8,\pi/4\}$ and $\mu=\{\mu_m\equiv\frac{1}{2}\sqrt{-3+\sqrt{17}},1,2,5,10,100\}$. The value of $\mu=\mu_m$ is the one for which the concurrence in the case of $RR$ incoming particles ($\eta=0$) is maximal. For this case we reproduce the results given in Refs.\cite{Cervera-Lierta:2017tdt,CerveraLierta:2019ejf}. The other values of $\mu$ have been chosen to show the entanglement behaviour from the non-relativistic regime to the relativistic one ($\mu=1$ corresponds to the characteristic scale for $|\vec{p}|=m_e$).
On the other hand, $\eta=\pi/8$ and $\eta=\pi/4$ represent the cases in which the main features of the entanglement distribution for $\theta\in[0,2\pi]$ become evident. 

By increasing the momentum, for $\eta \neq 0$, the concurrence of our reference state shows a shift of its peak and an asymmetry with respect to $\theta=\pi$ emerges due to the interference terms in the associated density matrix. Moreover, from Figs.\ref{fig2a}-\ref{fig3c} we can see that the concurrence value for $\eta=0$ decreases and tends to zero in the relativistic limit, while for the other two cases $\eta=\pi/8$ and $\eta=\pi/4$, it increases and in the relativistic limit stabilizes its maxima at $\theta=\pi/2$ and $\theta = 3\pi/2$. These two values of $\eta$ correspond to a change in the entanglement weight towards the case of $\eta=\pi/2$. In fact, by setting $\eta=\pi/2$ (see Fig.\ref{figfour}), we recover the other case in Refs.\cite{Cervera-Lierta:2017tdt,CerveraLierta:2019ejf} with $R$ and $L$ helicities for the incoming particles in which in the high energy limit maximal entanglement emerges in $\theta=\pi/2$ and $\theta=3\pi/2$. In the same limit,  the other cases ($\eta\neq 0,\pi/2$) show that the asymmetry in concurrence with respect to $\theta=\pi$ is suppressed.

We remark that this asymmetry, as shown in Fig.\ref{figfive}, is due to the choice of initial polarizations: for different incoming helicity states, the amplitudes $\mathcal{M}(RR,rs)$ and $\mathcal{M}(RL,rs)$ carry different weights to the final concurrence. Choosing the opposite polarizations for $A$ and $B$, we recover the reflected plot with respect to $\theta=\pi$.


%\newpage
\mbox{}
\clearpage

\section{Entanglement distribution in bipartite subsystems}

In Section III we have studied an extension of the setup considered in Refs.\cite{Cervera-Lierta:2017tdt, CerveraLierta:2019ejf, Fedida:2022izl}, in which we have considered the positron $B$ in a superposition of helicity states as  in Eq.\eqref{inrefstate}. 
Now we come back to the original system, where a spectator particle $C$ (electron) is entangled with the positron $B$ which scatters with the electron $A$ as represented pictorially in Fig.\ref{fig scheme}. The states before and after the scattering are expressed by: 
 
\begin{equation}
\label{instateel}
    \ket{i} = \ket{R}_A\otimes\Big(\cos{\eta}\ket{R}_B\ket{R}_C+e^{i\beta}\sin{\eta}\ket{L}_B\ket{L}_C\Big),
\end{equation}
and
\begin{equation}
\label{finstateel}
    \ket{f} = \sum\limits_{r,s=R,L}\Big[\cos{\eta}\mathcal{M}(RR;rs)\ket{r}_A\ket{s}_B\ket{R}_C +e^{i\beta}\sin{\eta}\mathcal{M}(RL;rs)\ket{r}_A\ket{s}_B\ket{L}_C\Big].
\end{equation}

From Eqs.\eqref{instateel} and \eqref{finstateel} we can calculate the density matrices of the system. We obtain (omitting normalization factors):
%that we obtain by tracing (see Eq.\eqref{trace}) with respect to $A$, $B$ and $C$:
\begin{eqnarray}
\label{rhoinstate}
 \nonumber
    \hspace{-2cm}\rho^{i}_{ABC} &= \cos^2{\eta}\,\ket{R}_A\ket{R}_B\ket{R}_C\prescript{}{C}{\bra{R}}\prescript{}{B}{\bra{R}\prescript{}{A}{\bra{R}}}
     \nonumber
    +e^{-i\beta}\sin{\eta}\cos{\eta}\,\ket{R}_A\ket{R}_B\ket{R}_C\prescript{}{C}{\bra{L}}\prescript{}{B}{\bra{L}\prescript{}{A}{\bra{R}}}
    \\ &+ e^{i\beta} \sin{\eta}\cos{\eta}\,\ket{R}_A\ket{L}_B\ket{L}_C\prescript{}{C}{\bra{R}}\prescript{}{B}{\bra{R}\prescript{}{A}{\bra{R}}}
    +\sin^2{\eta}\,\ket{R}_A\ket{L}_B\ket{L}_C\prescript{}{C}{\bra{L}}\prescript{}{B}{\bra{L}\prescript{}{A}{\bra{R}}} ,
\end{eqnarray}
\begin{eqnarray}
\label{rhofinstate}
 \nonumber
    \hspace{-2cm}\rho^{f}_{ABC} &= \sum\limits_{r,s,r^{{}_{'}},s^{{}_{'}}}&\!\!\!\Big[\cos^2{\eta}\,\mathcal{M}(RR,rs)\mathcal{M}^{\dagger}(RR,r^{{}_{'}}s^{{}_{'}})\ket{r}_A\ket{s}_B\ket{R}_C\prescript{}{C}{\bra{R}}\prescript{}{B}{\bra{s^{{}_{'}}}}\prescript{}{A}{\bra{r^{{}_{'}}}}
    \\ \nonumber
    &&+e^{-i\beta}\sin{\eta}\cos{\eta}\,\mathcal{M}(RR,rs)\mathcal{M}^{\dagger}(RL,r^{{}_{'}}s^{{}_{'}})\ket{r}_A\ket{s}_B\ket{R}_C\prescript{}{C}{\bra{L}}\prescript{}{B}{\bra{s^{{}_{'}}}}\prescript{}{A}{\bra{r^{{}_{'}}}}
    \\ \nonumber
    &&+e^{i\beta}\sin{\eta}\cos{\eta}\,\mathcal{M}(RL,rs)\mathcal{M}^{\dagger}(RR,r^{{}_{'}}s^{{}_{'}})\ket{r}_A\ket{s}_B\ket{L}_C\prescript{}{C}{\bra{R}}\prescript{}{B}{\bra{s^{{}_{'}}}}\prescript{}{A}{\bra{r^{{}_{'}}}}
    \\ 
    &&+\sin^2{\eta}\,\mathcal{M}(RL,rs)\mathcal{M}^{\dagger}(RL,r^{{}_{'}}s^{{}_{'}})\ket{r}_A\ket{s}_B\ket{L}_C\prescript{}{C}{\bra{L}}\prescript{}{B}{\bra{s^{{}_{'}}}}\prescript{}{A}\!{\bra{r^{{}_{'}}}} \Big].
\end{eqnarray}

By tracing with respect to $A,B,C$, the reduce density matrices result:
\begin{equation}
\label{rhoabin}
\hspace{-1cm}
    \rho^{i}_{AB} = \cos^2{\eta}\ket{R}_A\ket{R}_B\prescript{}{B}{\bra{R}}\prescript{}{A}{\bra{R}}+\sin^2{\eta}\ket{R}_A\ket{L}_B\prescript{}{B}{\bra{L}}\prescript{}{A}{\bra{R}},
\end{equation} 
\begin{equation}
\label{rhoacin}
    \rho^{i}_{AC} = \cos^2{\eta}\ket{R}_A\ket{R}_C\prescript{}{C}{\bra{R}}\prescript{}{A}{\bra{R}}+\sin^2{\eta}\ket{R}_A\ket{L}_C\prescript{}{C}{\bra{L}}\prescript{}{A}{\bra{R}},
\end{equation}
\begin{eqnarray}\nonumber
\label{rhobcin}
    &\rho^{i}_{BC} = &\!\!\!\cos^2{\eta}\ket{R}_B\ket{R}_C\prescript{}{C}{\bra{R}}\prescript{}{B}{\bra{R}}+e^{-i\beta}\sin{\eta}\cos{\eta}\ket{R}_B\ket{R}_C\prescript{}{C}{\bra{L}}\prescript{}{B}{\bra{L}}
    \\[2mm]
    &&     +e^{i\beta}\sin{\eta}\cos{\eta}\ket{L}_B\ket{L}_C\prescript{}{C}{\bra{R}}\prescript{}{B}{\bra{R}}+\sin^2{\eta}\ket{L}_B\ket{L}_C\prescript{}{C}{\bra{L}}\prescript{}{B}{\bra{L}}.
\end{eqnarray}

\smallskip

\begin{eqnarray}\nonumber
\label{rhoabfin}
\rho^{f}_{AB}&=\sum\limits_{r,s,r',s'}\Big[&\!\!\!\cos^2{\eta}\mathcal{M}(RR;rs)\mathcal{M^{\dagger}}(RR;r's')\ket{r}_A\ket{s}_B\prescript{}{B}{\bra{s'}}\prescript{}{A}{\bra{r'}}\\ 
&& +\sin^2{\eta}\mathcal{M}(RL;rs)\mathcal{M^{\dagger}}(RL;r's')\ket{r}_A\ket{s}_B\prescript{}{B}{\bra{s'}}\prescript{}{A}{\bra{r'}}\Big],
\end{eqnarray}
\begin{eqnarray}\nonumber
\label{rhoacfin}
    \rho^{f}_{AC} &= \sum\limits_{r,r',s}\Big[&\!\!\!\cos^2{\eta}\,\mathcal{M}(RR;rs)\mathcal{M^{\dagger}}(RR;r's)\ket{r}_A\ket{R}_C\prescript{}{C}{\bra{R}}\prescript{}{A}{\bra{r'}}
    \\ [2mm] \nonumber
    &&+e^{-i\beta}\sin{\eta}\cos{\eta}\,\mathcal{M}(RR;rs)\mathcal{M^{\dagger}}(RL;r's)\ket{r}_A\ket{R}_C\prescript{}{C}{\bra{L}}\prescript{}{A}{\bra{r'}} 
    \\[2mm] \nonumber
    && +e^{i\beta}\sin{\eta}\cos{\eta}\,\mathcal{M}(RL;rs)\mathcal{M^{\dagger}}(RR;r's)\ket{r}_A\ket{L}_C\prescript{}{C}{\bra{R}}\prescript{}{A}{\bra{r'}}
    \\[2mm] 
    &&+\sin^2{\eta}\,\mathcal{M}(RL;rs)\mathcal{M^{\dagger}}(RL;r's)\ket{r}_A\ket{L}_C\prescript{}{C}{\bra{L}}\prescript{}{A}{\bra{r'}}\Big],
\end{eqnarray}  
\begin{eqnarray}\nonumber
\label{rhobcfin}
    \rho^{f}_{BC}& = \sum\limits_{r,r,s'}&\Big[\cos^2{\eta}\,\mathcal{M}(RR;rs)\mathcal{M^{\dagger}}(RR;rs')\ket{s}_B\ket{R}_C\prescript{}{C}{\bra{R}}\prescript{}{B}{\bra{s'}}\\ [2mm] \nonumber
    &&+e^{-i\beta}\sin{\eta}\cos{\eta}\,\mathcal{M}(RR;rs)\mathcal{M^{\dagger}}(RL;rs')\ket{s}_B\ket{R}_C\prescript{}{C}{\bra{L}}\prescript{}{B}{\bra{s'}}
    \\ [2mm] \nonumber
    &&    +e^{i\beta}\sin{\eta}\cos{\eta}\,\mathcal{M}(RL;rs)\mathcal{M^{\dagger}}(RR;rs')\ket{s}_B\ket{L}_C\prescript{}{C}{\bra{R}}\prescript{}{B}{\bra{s'}}
    \\ [2mm]
    &&+\sin^2{\eta}\,\mathcal{M}(RL;rs)\mathcal{M^{\dagger}}(RL;rs')\ket{s}_B\ket{L}_C\prescript{}{C}{\bra{L}}\prescript{}{B}{\bra{s'}}\Big].
\end{eqnarray}

\smallskip
Using Eqs.\eqref{conc} and $\eqref{matconc}$ we calculate the concurrence associated to each of the six bipartite systems \eqref{rhoabin}-\eqref{rhobcfin}. As before, we take $\beta=0$. For the bipartitions of the initial state, we obtain $C(\rho^i_{AB})=0$, $C(\rho^i_{AC})=0$  and $C(\rho^i_{BC})=|\sin(2\eta)|$.  
On the other hand, the expressions of concurrence for the bipartitions of the final states  are very lengthy and they are not reported here except for those in the relativistic limit, that we list below:
\bea
\label{abfinlim}
    \lim_{\mu\rightarrow\infty}C(\rho^f_{AB})
    %&=&\frac{64\bigl(1-\cos(2\eta)\bigr)\sin^4(\theta/2)\cos^4(\theta/2)}{64\cos^2(\eta)+(35+28\cos(2\theta)+\cos(4\theta))\sin^2(\eta)}, \\
    &=&\frac{2 \sin^2\eta \sin^4(\theta/2)\cos^4(\theta/2)}{1-(1-\frac{1}{8}\sin^2\theta)\sin^2\theta\sin^2\eta},
\eea
\begin{equation}
\label{acfinlim}
    \lim_{\mu\rightarrow\infty}C(\rho^f_{AC})=
    \frac{\sin(2\eta)\sin^4(\theta/2)}{1-(1-\frac{1}{8}\sin^2\theta)\sin^2\theta\sin^2\eta},
   % \frac{64\sin(2\eta)\sin^4(\theta/2)}{64\cos^2(\eta)+(35+28\cos(2\theta)+\cos(4\theta))\sin^2(\eta)},
\end{equation}
\begin{equation}
\label{bcfinlim}
    \lim_{\mu\rightarrow\infty}C(\rho^f_{BC})=
    \frac{\sin(2\eta)\cos^4(\theta/2)}{1-(1-\frac{1}{8}\sin^2\theta)\sin^2\theta\sin^2\eta}.
    %\frac{64\sin(2\eta)\cos^4(\theta/2)}{64\cos^2(\eta)+(35+28\cos(2\theta)+\cos(4\theta))\sin^2(\eta)}.
\end{equation}

\begin{figure}[h]
 \subfloat[][\emph{}]{\includegraphics[width =5.55 cm]{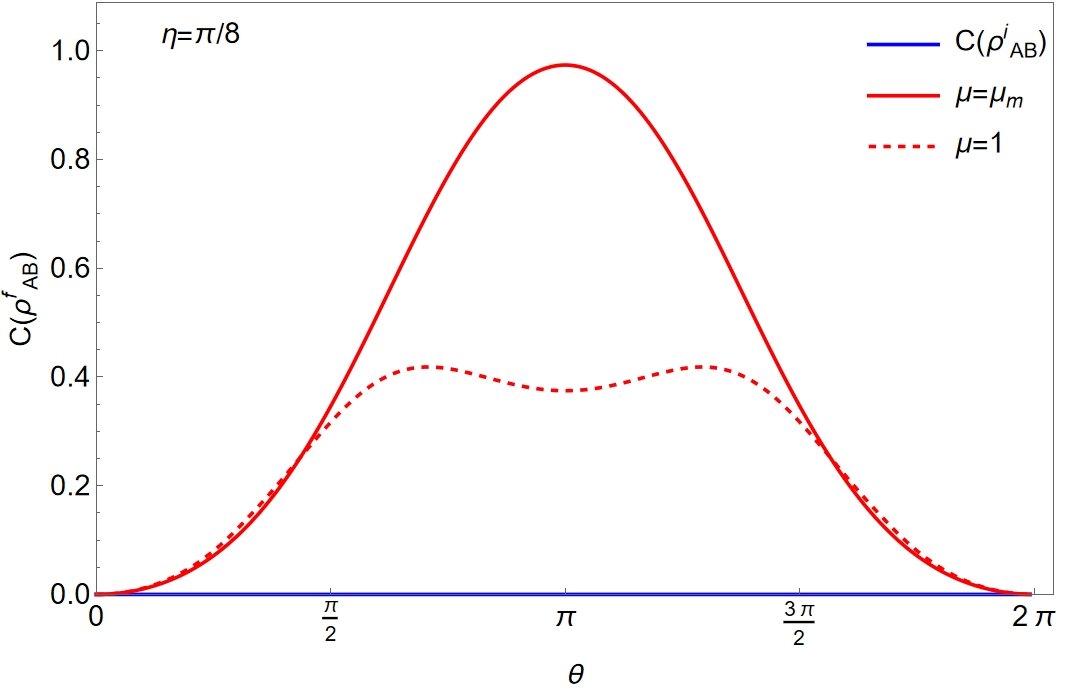}
\label{fig6a}}\quad
 \subfloat[][\emph{}]{\includegraphics[width =5.55 cm]{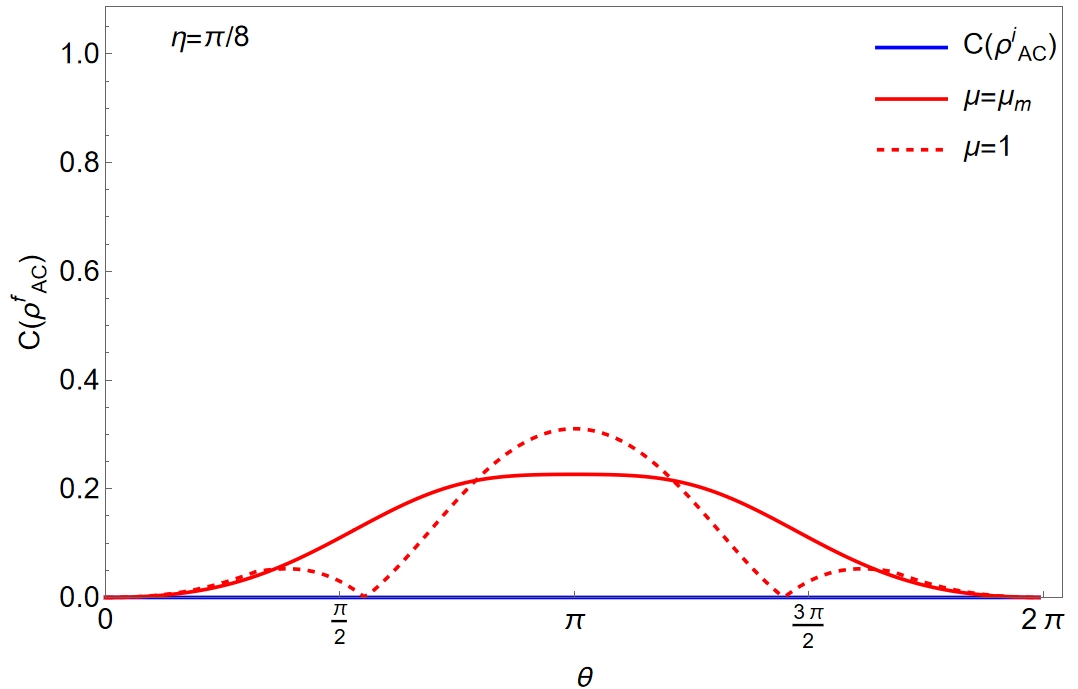}
\label{fig6b}}\quad
 \subfloat[][\emph{}]{\includegraphics[width =5.55 cm]{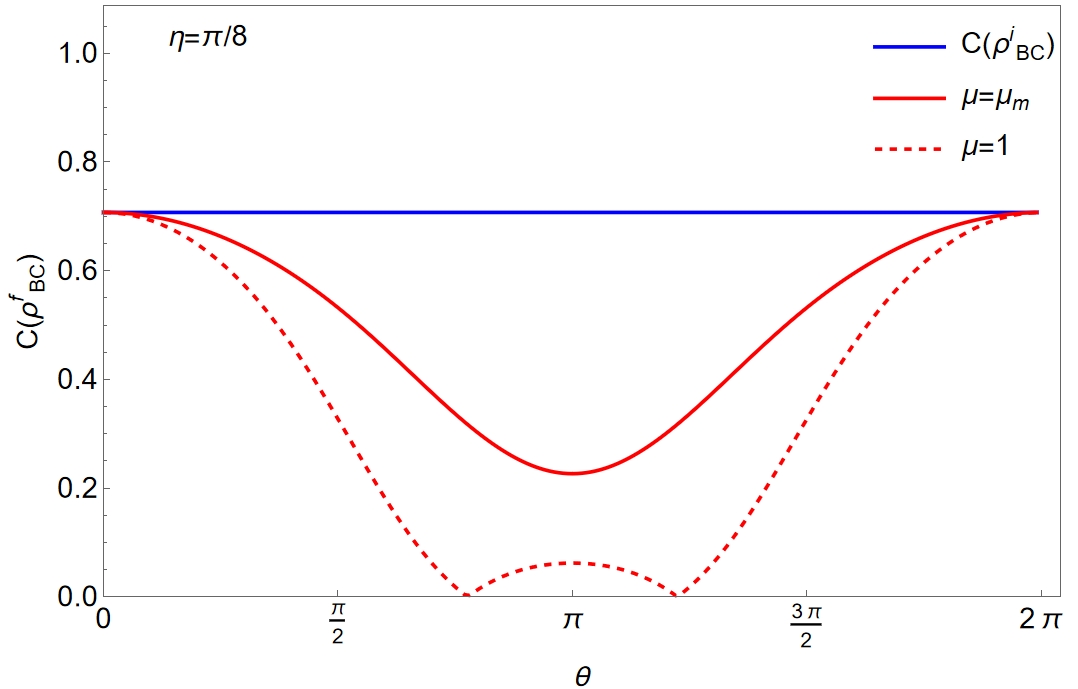}
\label{fig6c}}\quad
 \subfloat[][\emph{}]{\includegraphics[width =5.55 cm]{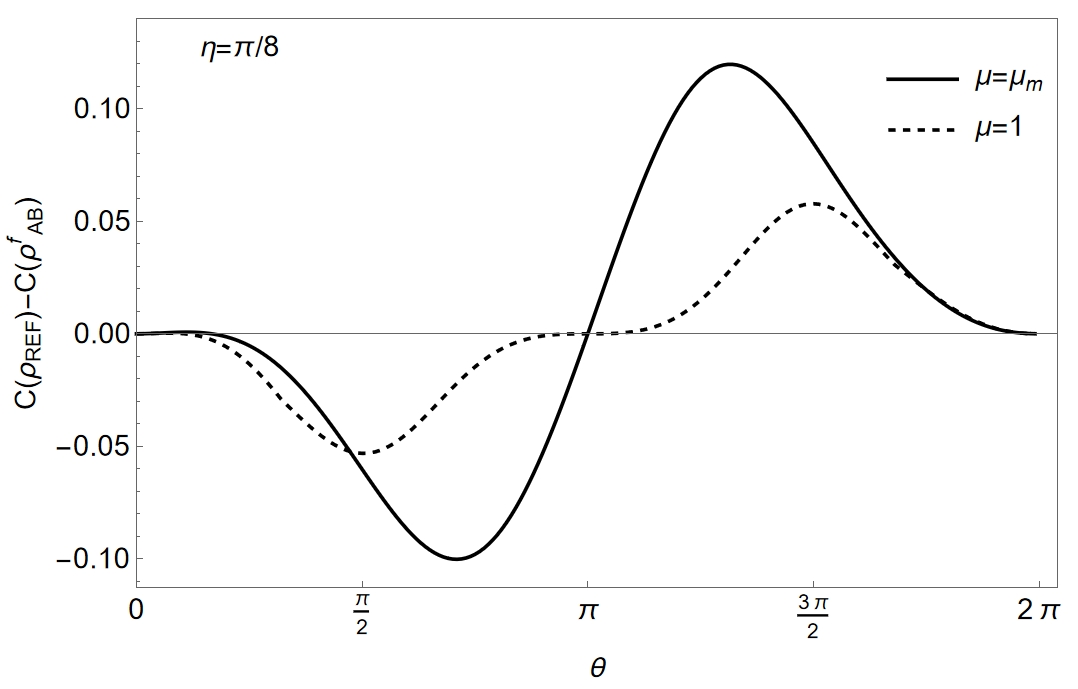}
\label{fig6d}}\quad
\caption{Concurrence of the three final bipartite systems: channel $AB$ Eq.\eqref{rhoabfin} (a), channel $AC$ Eq. \eqref{rhoacfin} (b), and channel $BC$ Eq. \eqref{rhobcfin} (c). In (d) the concurrence difference $C(\rho_{REF})-C(\rho_{AB})$.} 
\label{figsix}
\end{figure}

\begin{figure}[!]
 \subfloat[][\emph{}]{\includegraphics[width =5.55 cm]{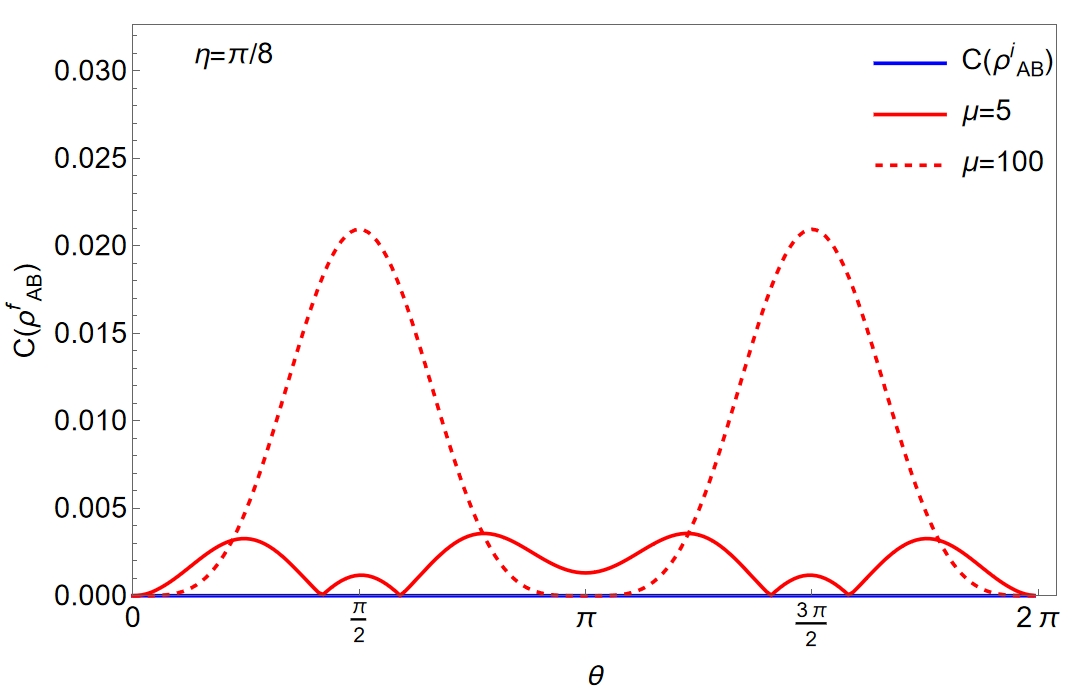}
\label{fig7a}}\quad
 \subfloat[][\emph{}]{\includegraphics[width =5.55 cm]{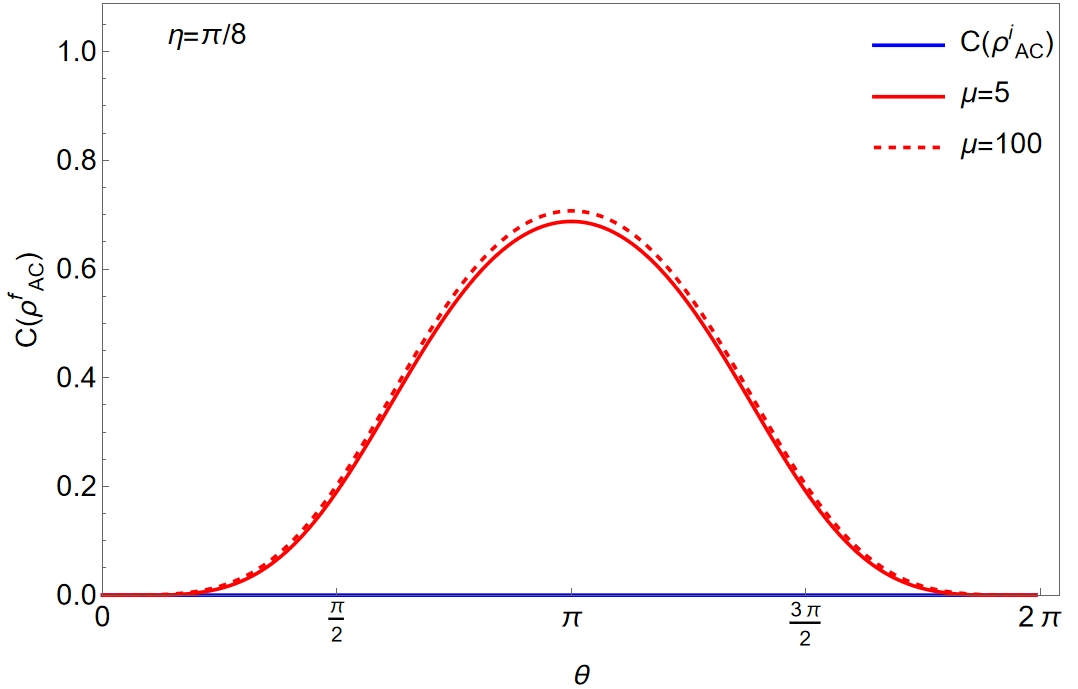}
\label{fig7b}}\quad
 \subfloat[][\emph{}]{\includegraphics[width =5.55 cm]{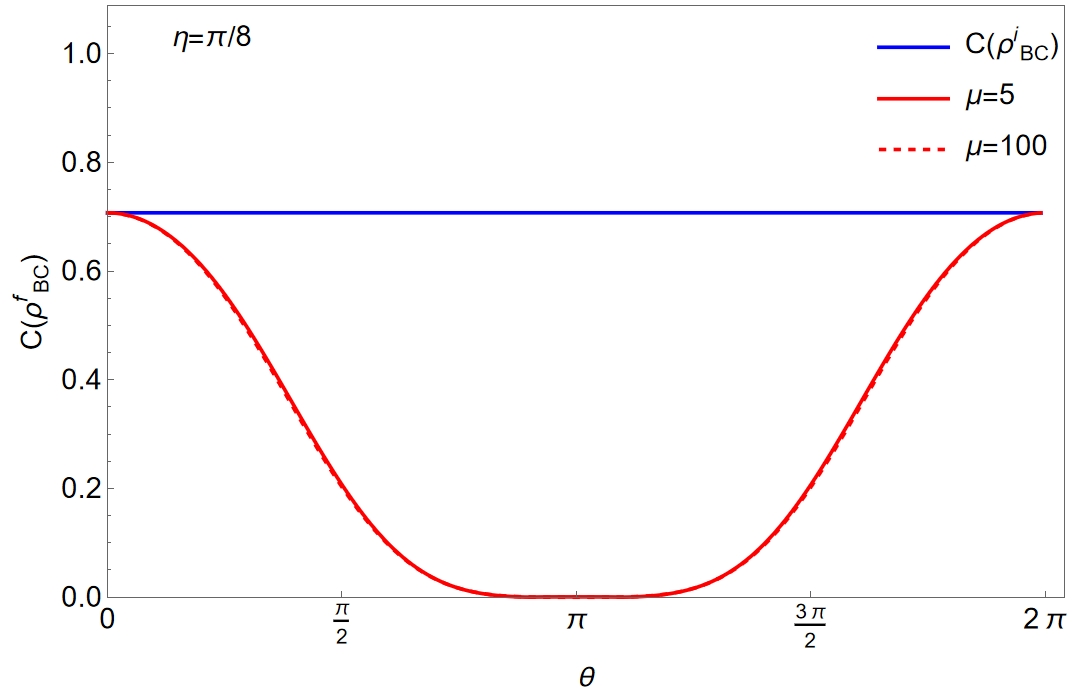}
\label{fig7c}}\quad
 \subfloat[][\emph{}]{\includegraphics[width =5.55 cm]{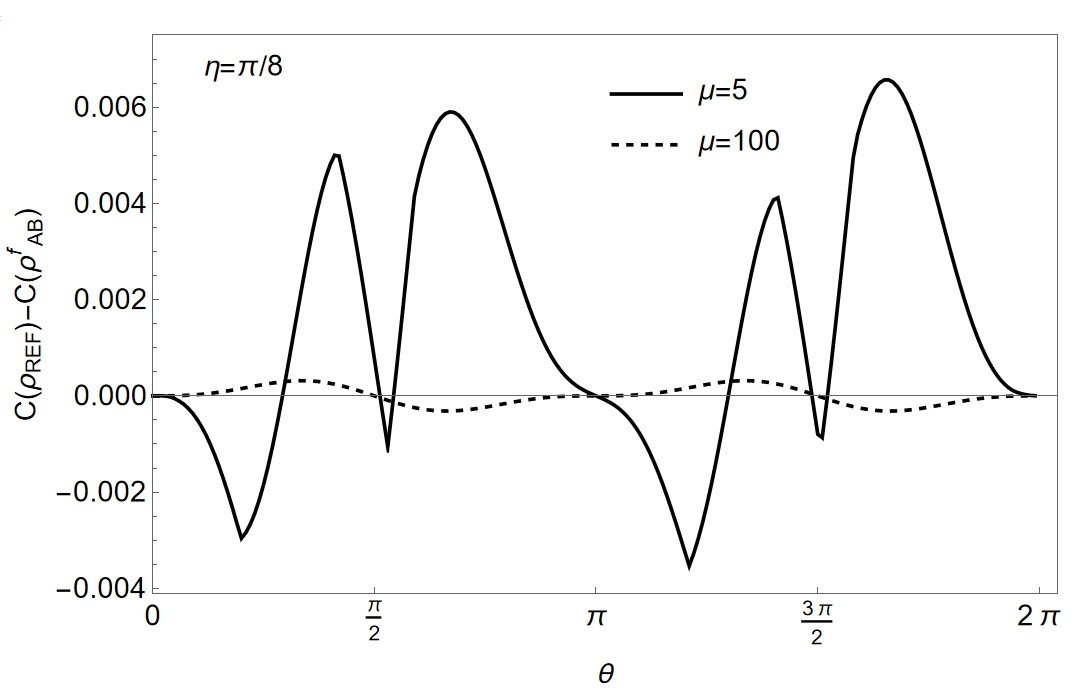}
\label{fig7d}}\quad
\caption{Concurrence of the three final bipartite systems: channel $AB$ Eq.\eqref{rhoabfin} (a), channel $AC$ Eq. \eqref{rhoacfin} (b), and channel $BC$ Eq. \eqref{rhobcfin} (c). In (d) the concurrence difference $C(\rho_{REF})-C(\rho_{AB})$.} 
\label{figseven}
\end{figure}

\begin{figure}
 \subfloat[][\emph{}]{\includegraphics[width =5.55 cm]{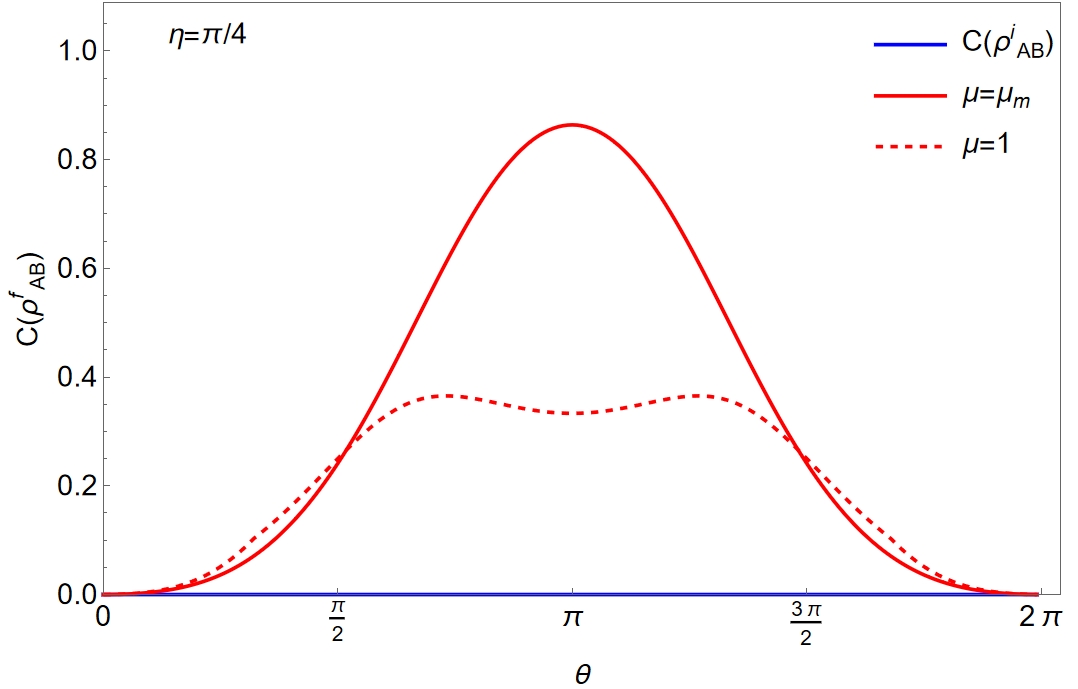}
\label{fig8a}}\quad
 \subfloat[][\emph{}]{\includegraphics[width =5.55 cm]{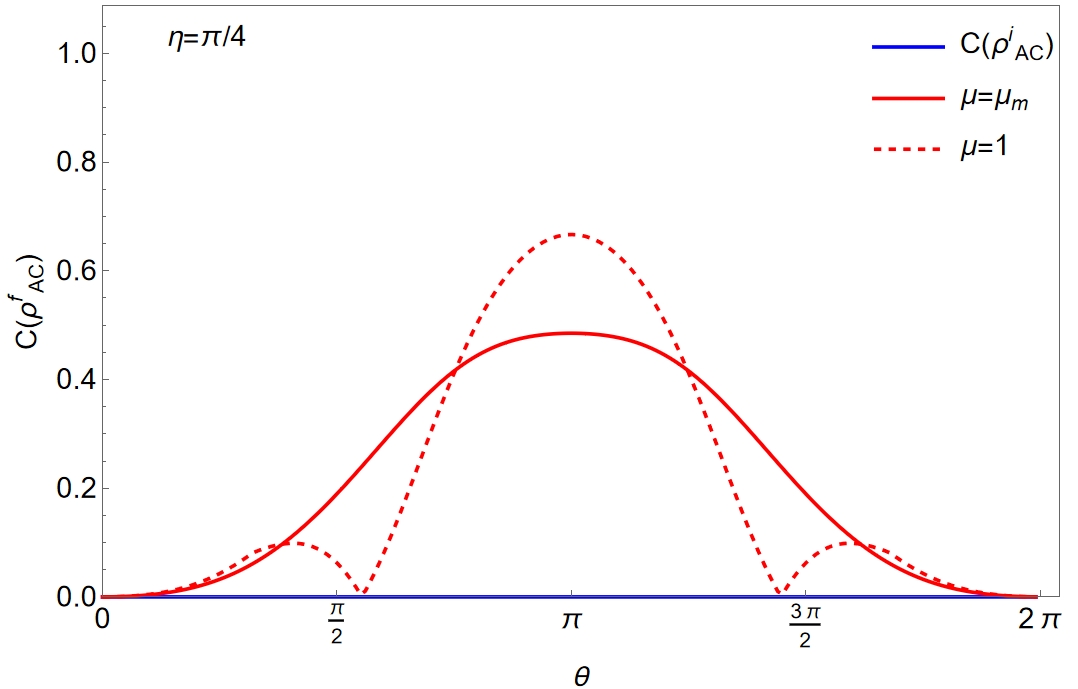}
\label{fig8b}}\quad
 \subfloat[][\emph{}]{\includegraphics[width =5.55 cm]{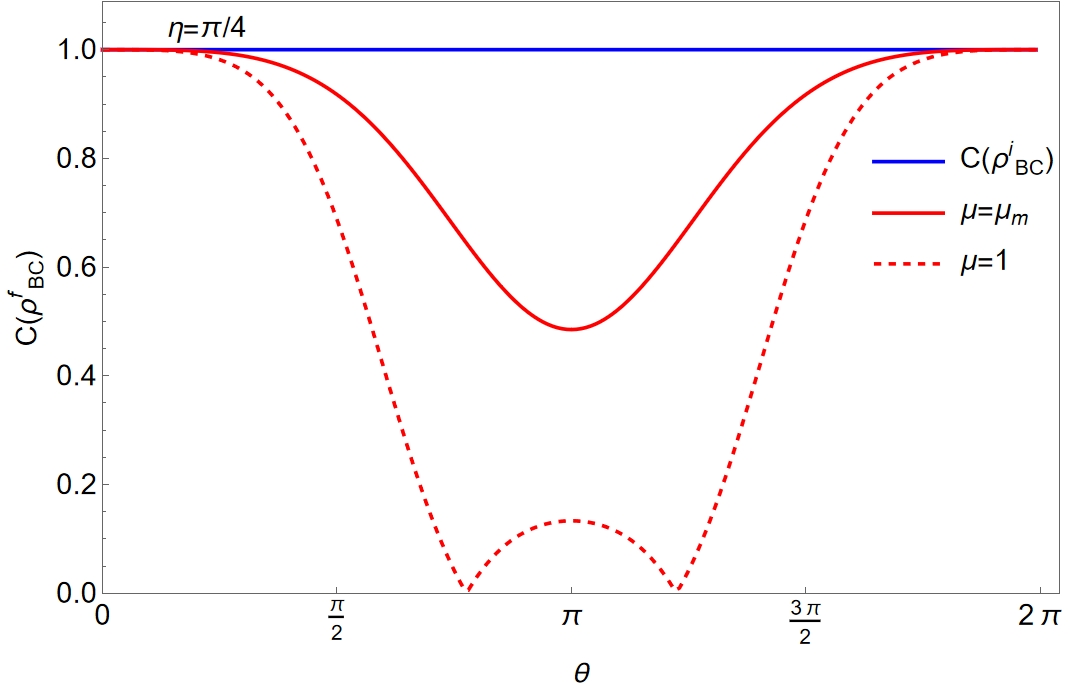}
\label{fig8c}}\quad
 \subfloat[][\emph{}]{\includegraphics[width =5.55 cm]{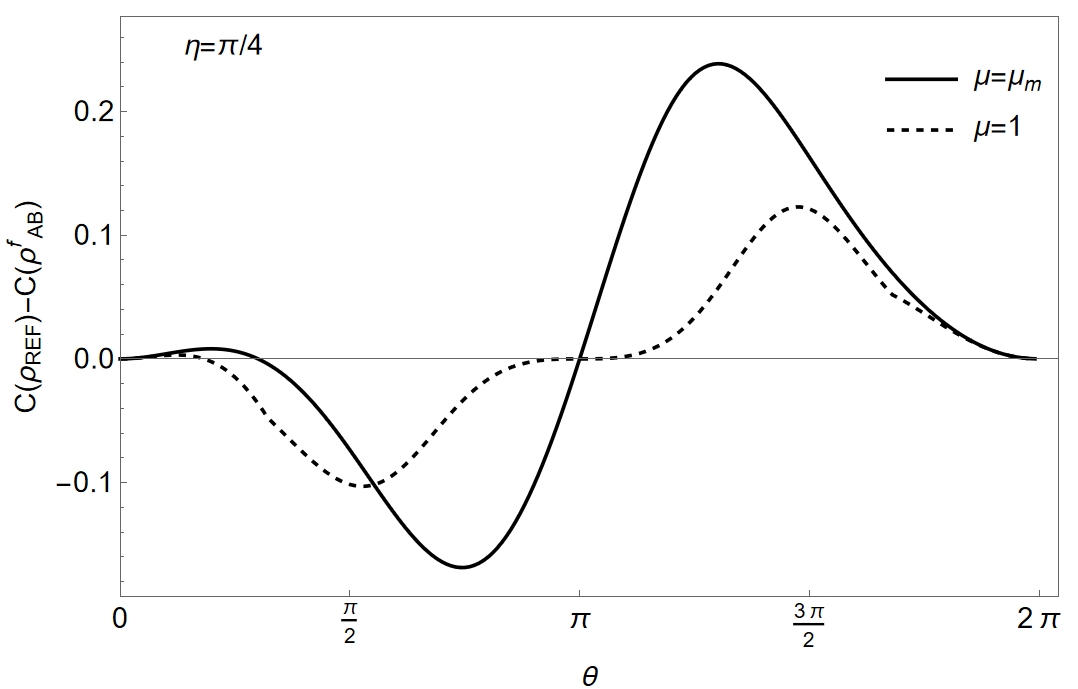}
\label{fig8d}}\quad
\caption{Concurrence of the three final bipartite systems: channel $AB$ Eq.\eqref{rhoabfin} (a), channel $AC$ Eq. \eqref{rhoacfin} (b), and channel $BC$ Eq. \eqref{rhobcfin} (c). In (d) the concurrence difference $C(\rho_{REF})-C(\rho_{AB})$.} 
\label{figeight}
\end{figure}

\begin{figure}
 \subfloat[][\emph{}]{\includegraphics[width =5.55 cm]{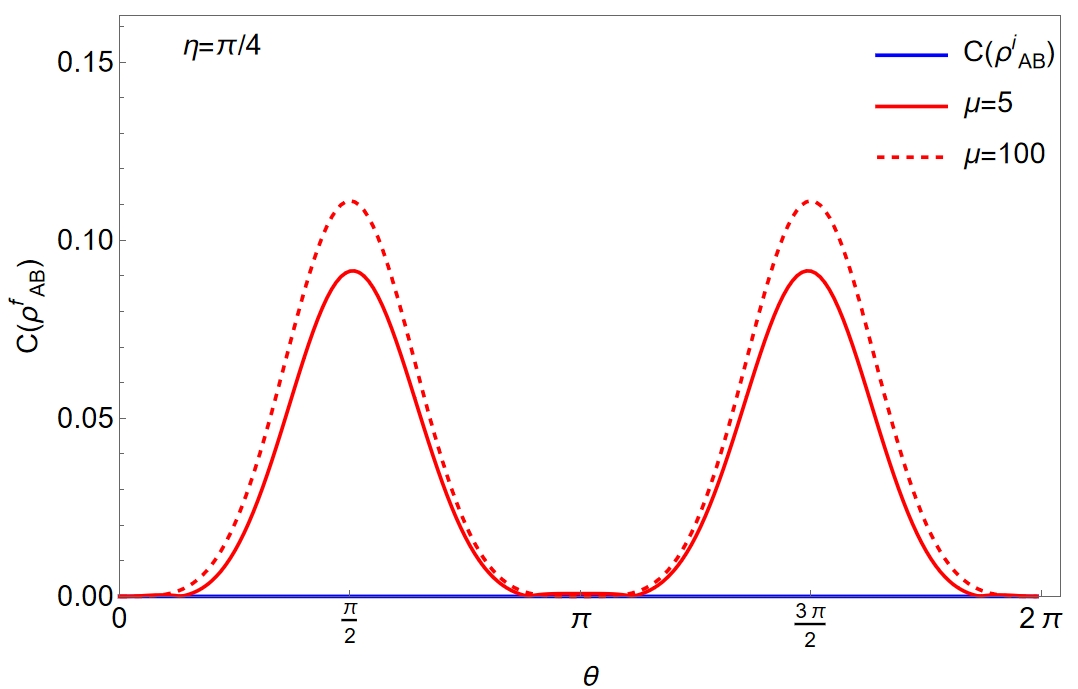}
\label{fig9a}}\quad
 \subfloat[][\emph{}]{\includegraphics[width =5.55 cm]{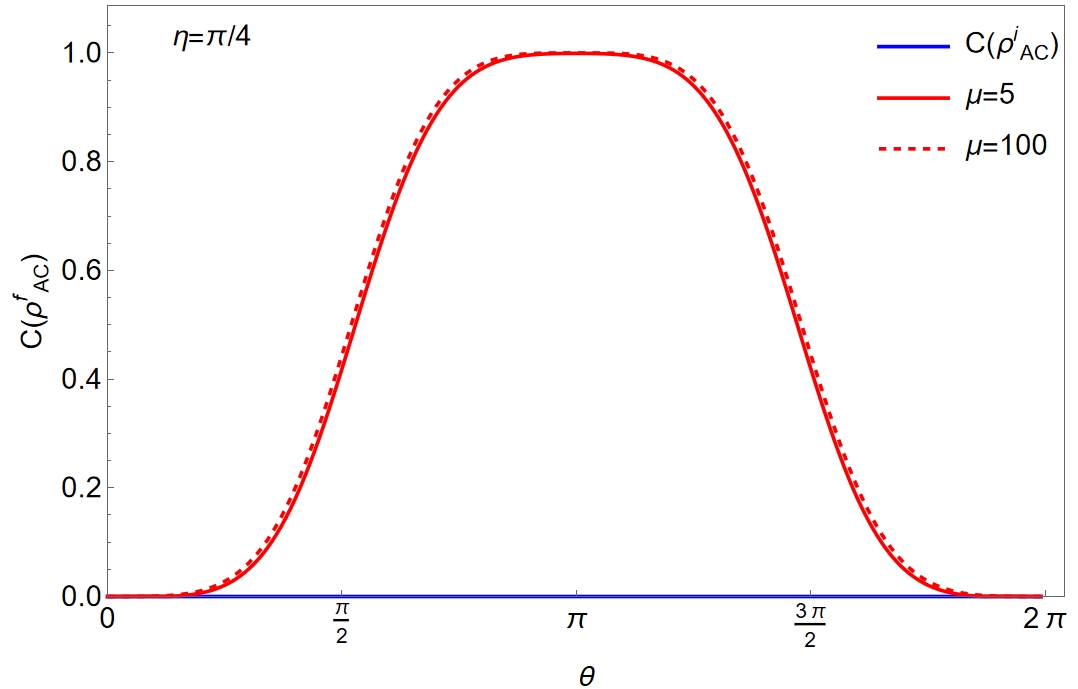}
\label{fig9b}}\quad
 \subfloat[][\emph{}]{\includegraphics[width =5.55 cm]{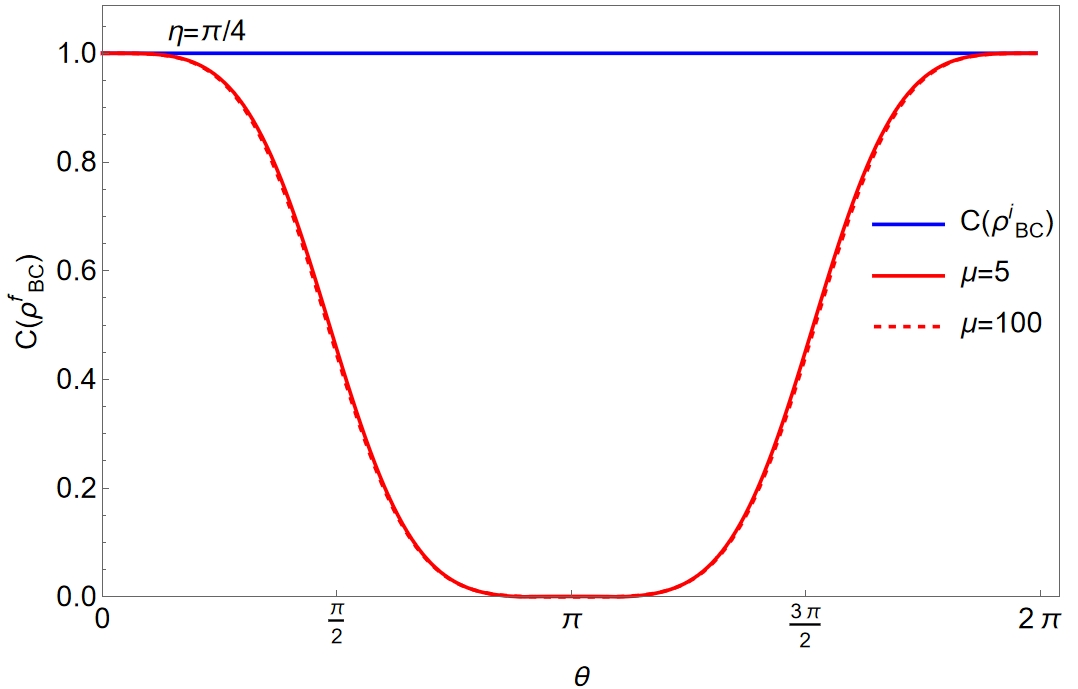}
\label{fig9c}}\quad
 \subfloat[][\emph{}]{\includegraphics[width =5.55 cm]{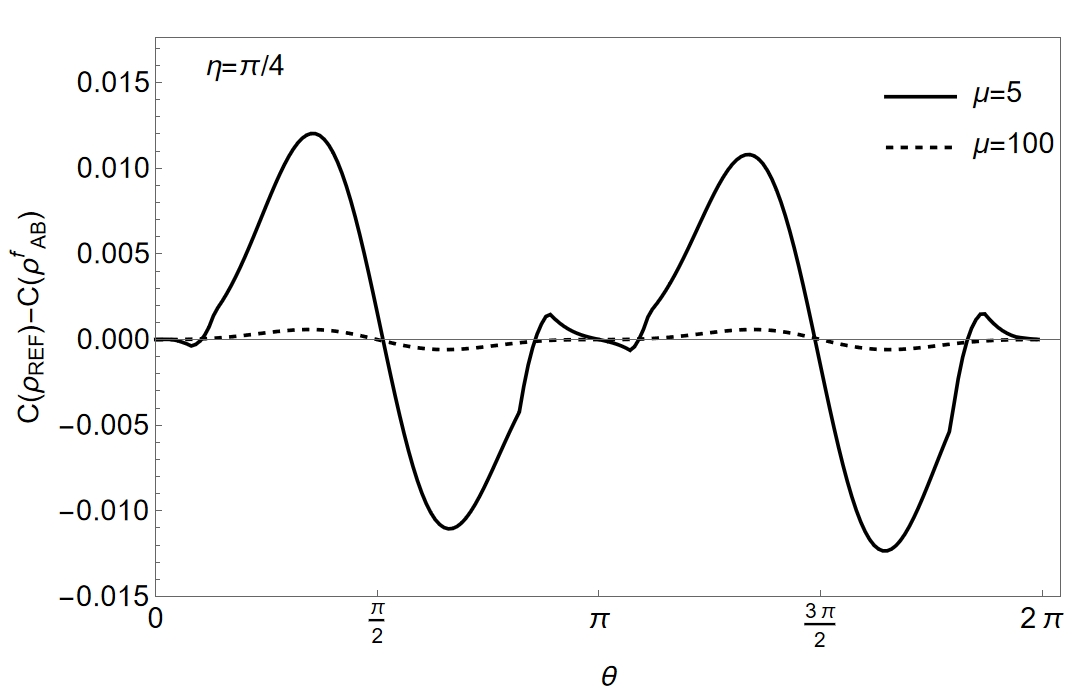}
\label{fig9d}}\quad
\caption{Concurrence of the three final bipartite systems: channel $AB$ Eq.\eqref{rhoabfin} (a), channel $AC$ Eq. \eqref{rhoacfin} (b), and channel $BC$ Eq. \eqref{rhobcfin} (c). In (d) the concurrence difference $C(\rho_{REF})-C(\rho_{AB})$.} 
\label{fignine}
\end{figure}

\begin{figure}
 \subfloat[][\emph{}]{\includegraphics[width =5.55 cm]{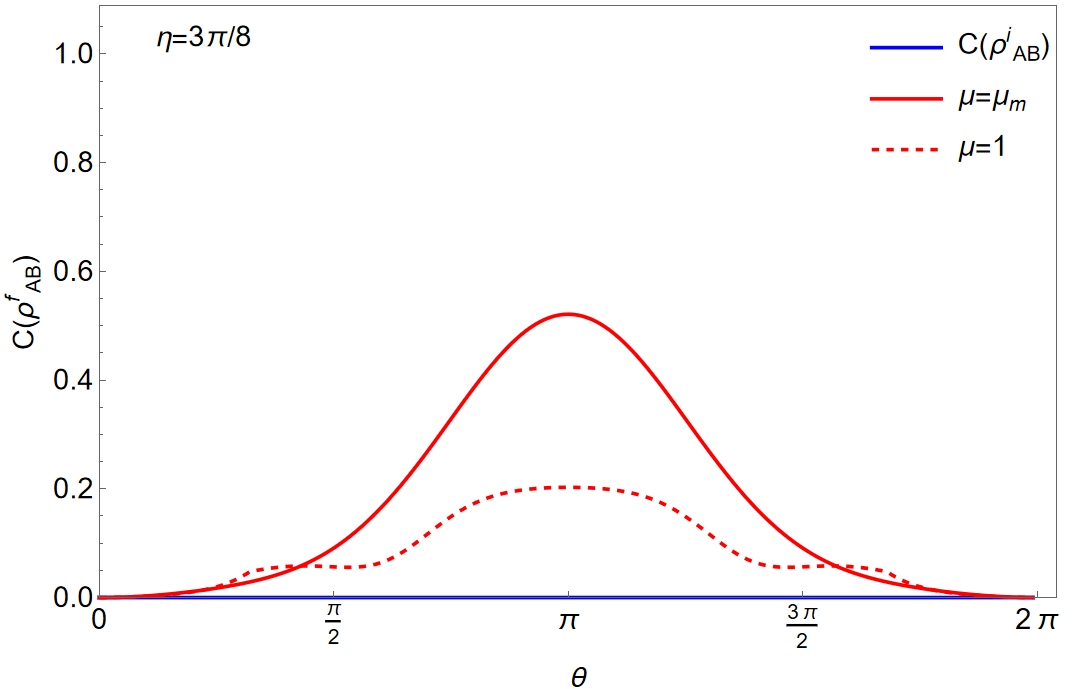}
\label{fig10a}}\quad
 \subfloat[][\emph{}]{\includegraphics[width =5.55 cm]{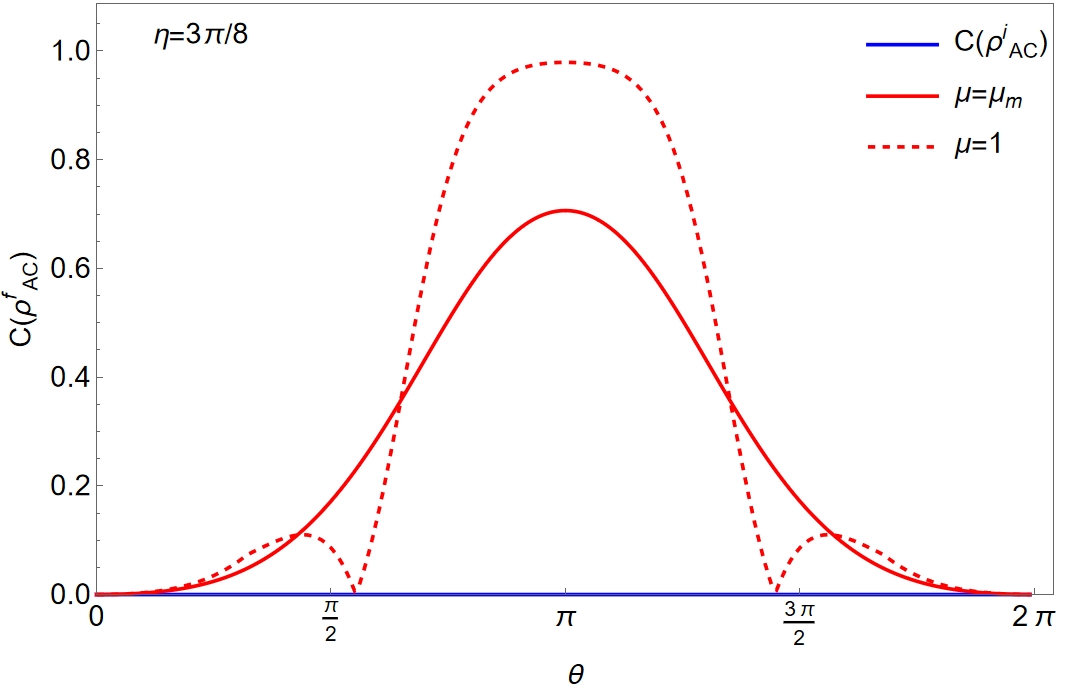}
\label{fig10b}}\quad
 \subfloat[][\emph{}]{\includegraphics[width =5.55 cm]{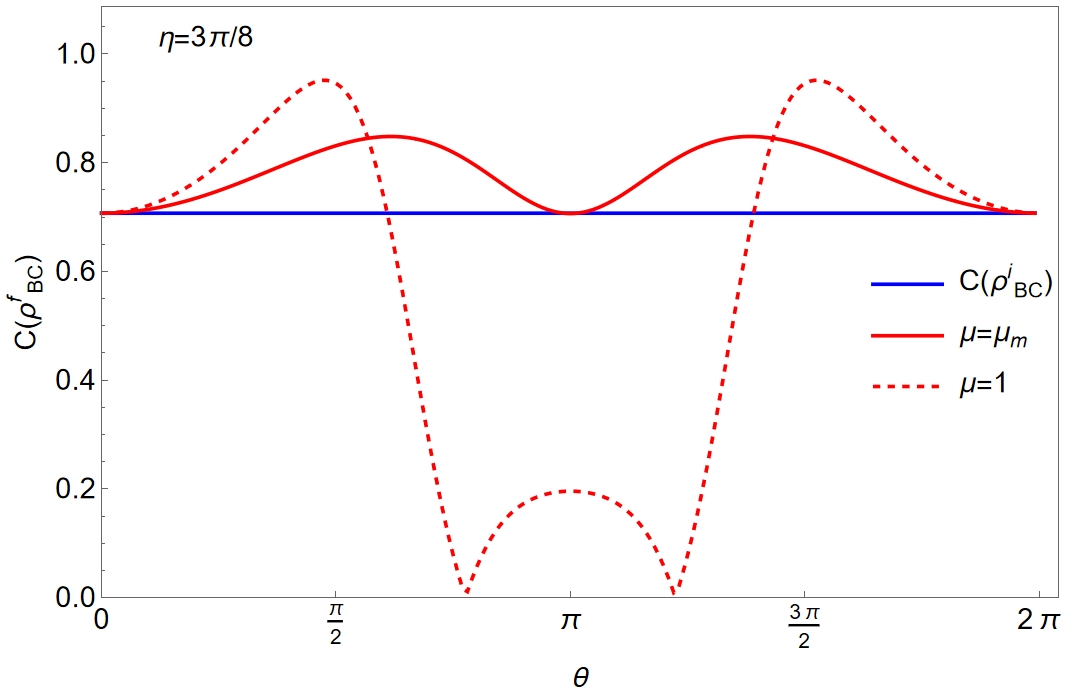}
\label{fig10c}}\quad
 \subfloat[][\emph{}]{\includegraphics[width =5.55 cm]{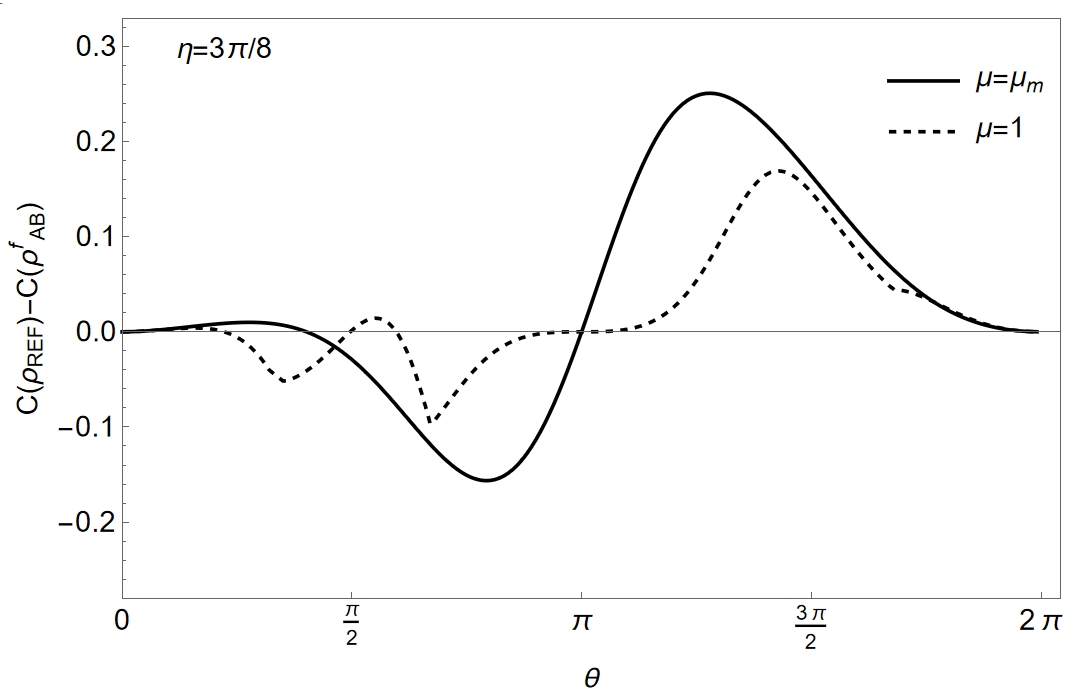}
\label{fig10d}}\quad
\caption{Concurrence of the three final bipartite systems: channel $AB$ Eq.\eqref{rhoabfin} (a), channel $AC$ Eq. \eqref{rhoacfin} (b), and channel $BC$ Eq. \eqref{rhobcfin} (c). In (d) the concurrence difference $C(\rho_{REF})-C(\rho_{AB})$.} 
\label{figten}
\end{figure}

\begin{figure}
 \subfloat[][\emph{}]{\includegraphics[width =5.55 cm]{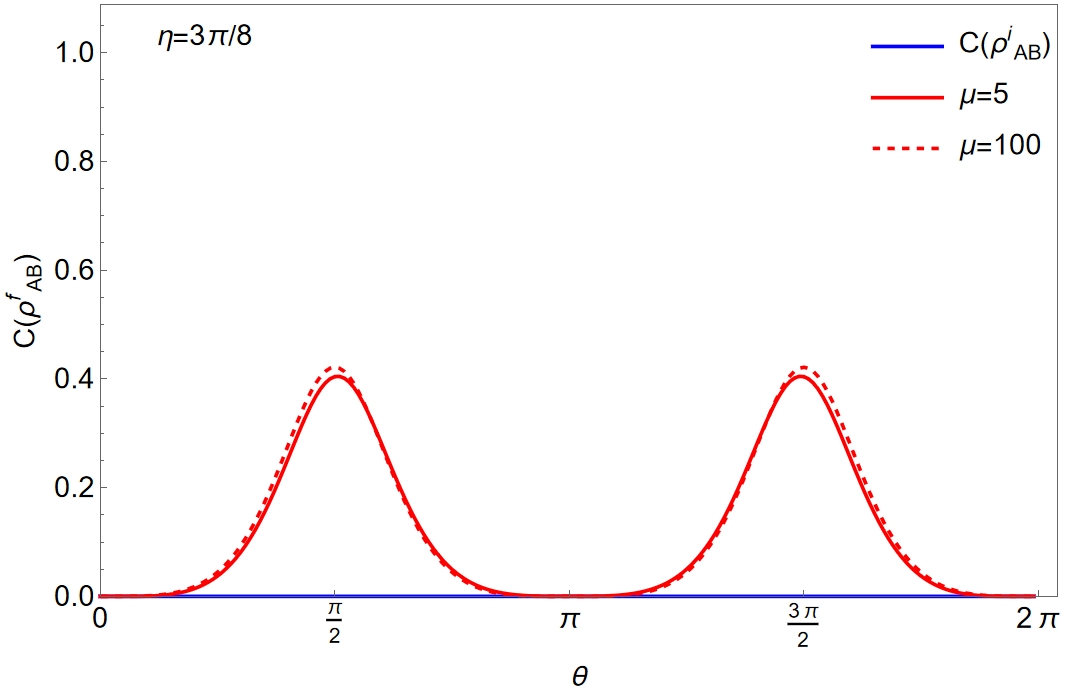}
\label{fig11a}}\quad
 \subfloat[][\emph{}]{\includegraphics[width =5.55 cm]{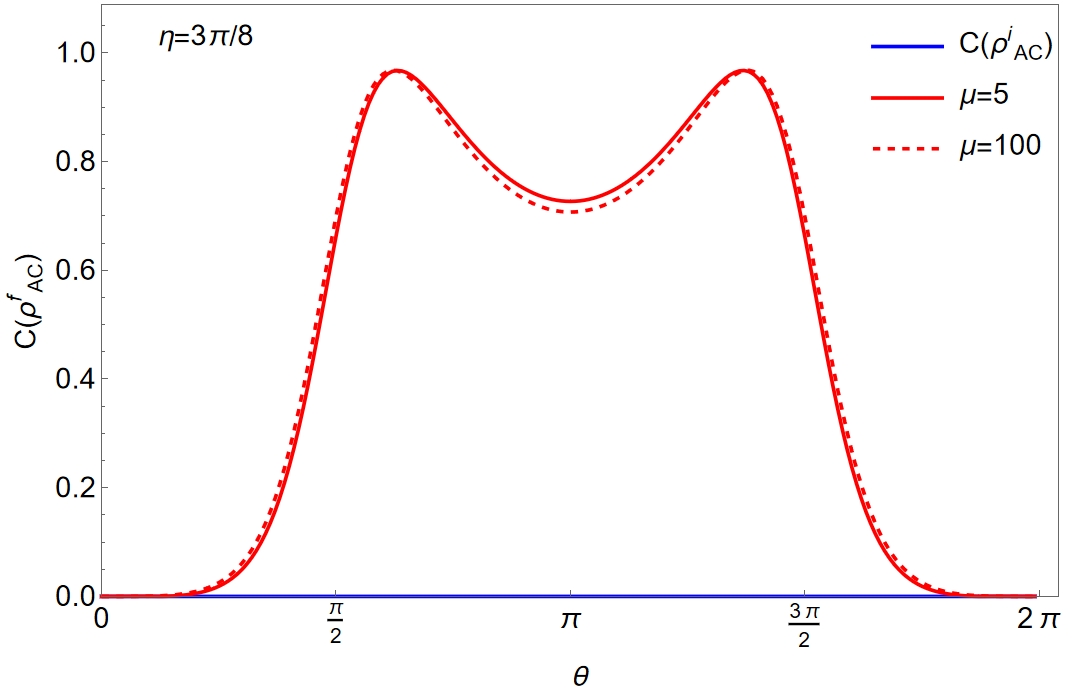}
\label{fig11b}}\quad
 \subfloat[][\emph{}]{\includegraphics[width =5.55 cm]{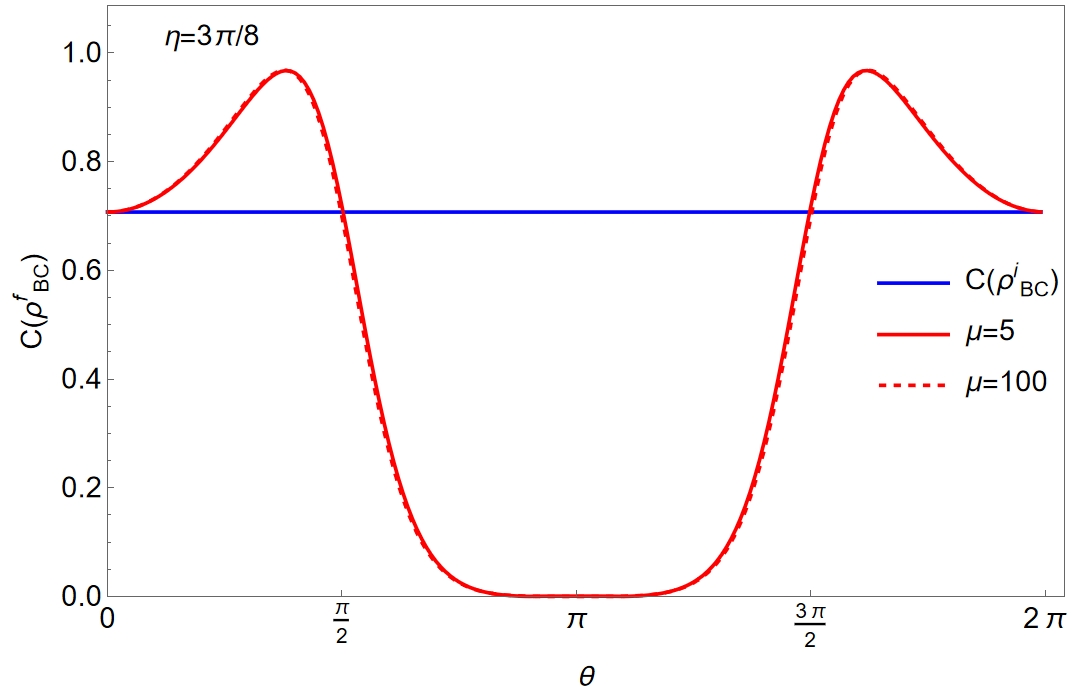}
\label{fig11c}}\quad
 \subfloat[][\emph{}]{\includegraphics[width =5.55 cm]{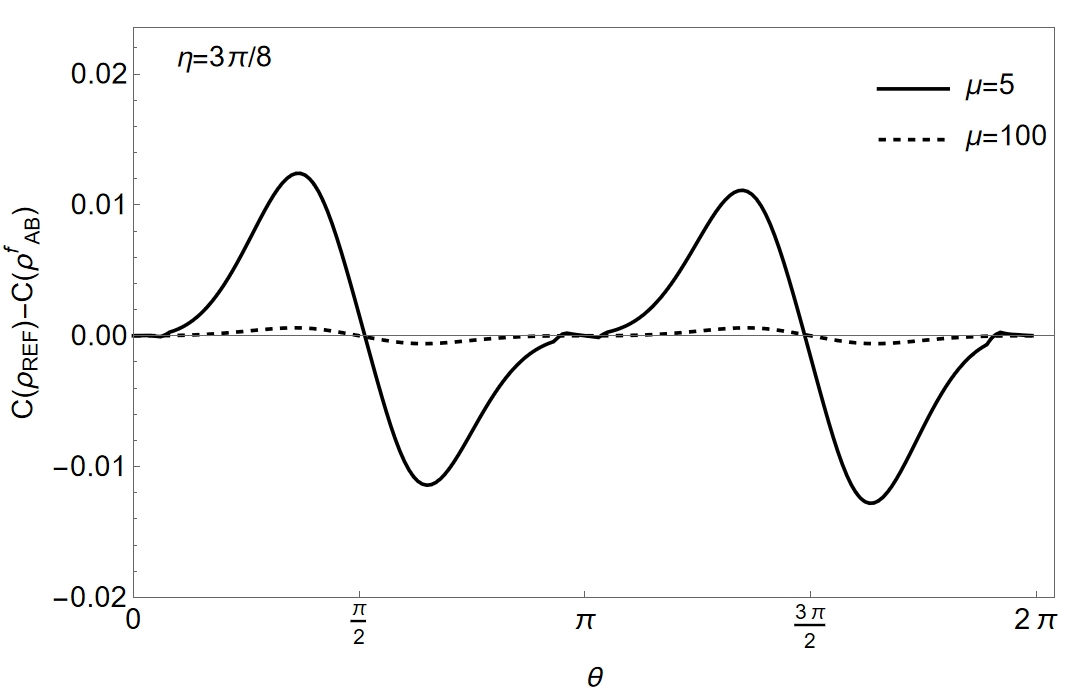}
\label{fig11d}}\quad
\caption{Concurrence of the three final bipartite systems: channel $AB$ Eq.\eqref{rhoabfin} (a), channel $AC$ Eq. \eqref{rhoacfin} (b), and channel $BC$ Eq. \eqref{rhobcfin} (c). In (d) the concurrence difference $C(\rho_{REF})-C(\rho_{AB})$.} 
\label{figeleven}
\end{figure}

Some representative plots of concurrences numerically evaluated are reported in Figs.\ref{figsix}-\ref{figeleven}.
For each set of plots, we have chosen four values of  $\mu$:$\{\mu_m, 1, 5, 100\}$ and  $\eta$:$\{0, \pi/8, \pi/4, 3\pi/8\}$, which are sufficient to represent clearly the generation and distribution of the entanglement in the three channels $AB$, $AC$ and $BC$. The plots (a), (b) and (c) in each figure represent the concurrence as a function of the scattering angle $\theta$ and  (d) the concurrence difference in the $AB$ channel between the reference state and the system.

In each set corresponding to $\eta=\pi/8$ and $\eta=\pi/4$ from Fig.\ref{figsix} to Fig.\ref{fignine}, the concurrence in the $BC$ channel decreases with respect to its initial value while in the $AC$ channel increases. 
On the other hand, in the same range of parameters, concurrence in the $AB$  channel has a nontrivial behaviour, which however stabilizes in the relativistic limit for $\theta=\frac{1}{2}\pi$ and $\theta=\frac{3}{2}\pi$, where it is maximal as for the (initial) disentangled case $\eta=\pi/2$ (i.e. with $RL$ incoming polarization particles) as reported in  \cite{Cervera-Lierta:2017tdt} and in Section III. We find remarkable that the correlation between the two particles $A$ and $C$ that not interact directly increases as a consequence of the scattering between $A$ and $B$.

On the other hand, when we consider the cases for $\eta>\pi/4$ up to $\eta=\frac{3}{4}\pi$,  entanglement in the $BC$ channel may also increases with respect to its initial value. This is clear from Figs.\ref{figten} and Fig.\ref{figeleven} for the particular case $\eta=\frac{3}{8}\pi$.

\medskip

Let us consider now the relativistic limit, in which the analytic expressions Eqs.\eqref{abfinlim}-\eqref{bcfinlim} for the concurrences are available. We observe that in such a limit, the entanglement in the $AB$ channel Eq.\eqref{abfinlim} is identical to the one for the reference state Eq.\eqref{ConcRefRel}: thus it appears to be completely generated in the scattering process.

We now further specialize to  the particular case of $\eta=\pi/4$, corresponding to maximal entanglement for the initial state $BC$.
From Eq.\eqref{acfinlim}, we see that at $\theta=\pi$, the entanglement in the $AC$ output channel is maximal, while it vanishes in the $BC$ output channel. Thus the QED scattering between $A$ and $B$ acts as a quantum gate for the complete transfer of the entanglement between $AC$ and $BC$. 

We also notice that correspondingly, the entanglement generated in the scattering assumes low values, thus entanglement transfer is a dominant and stable mechanism around $\theta=\pi$ in the relativistic limit and for entanglement weight $\eta=\pi/4$.

\newpage

\clearpage

\section{CONCLUSIONS AND OUTLOOK}
In this work, we have studied entanglement in the context of QED processes. In particular, we have considered  Bhabha scattering at tree-level in which a positron $B$, that scatters with an electron $A$, is entangled in spin with another electron $C$ that does not participate directly in the process. 
We found that two effects occur: the entanglement generation at the interaction vertex and the distribution of the initial entanglement among the three channels. 

By using the concurrence, we have quantified the entanglement in the three bipartitions of the system: $AB$, $AC$ and $BC$ -- before and after the scattering. The correlations depend on the value of the entanglement weight $\eta$, the scattering angle $\theta$, and the ratio between the incoming momentum and the mass $\mu$. The interplay between generation and transfer of entanglement among the three channels is very complex in the non-relativistic regime, namely for $\mu\simeq 1$. On the other hand, for some configurations of parameters, the entanglement tends to concentrate in some bipartitions.

Especially interesting is the relativistic regime $\mu=\infty$. In such a limit, we were able to calculate the analytic expressions for the concurrences and found that the entanglement in the $AB$ output channel is not affected by the presence of the entangled spectator particle, thus being completely generated in the scattering. 
On the other hand, for $\eta=\pi/4$, we observe a complete transfer of entanglement from the $BC$ channel to the $AC$ channel in a neighbour of $\theta=\pi$. In this situation, the QED scattering between $A$ and $B$ acts as a quantum gate for such a transfer between $AC$ and $BC$. It is an intriguing question if such a mechanism could be useful for quantum information tasks.

%Also, it is very interesting to note that concurrence for our reference state, in which only $A$ and $B$ particle are involved in the process, is antisymmetric with respect to $\theta = \pi$ (of course with the opposite chose in polarization the behaviour in concurrence is the same reflected with respect to $\theta = \pi$ as shown in Fig.\ref{figfive}). This property vanishes if it is considered the scattering with $B$ and $C$ entangled (the concurrence for $AB$, $AC$ and $BC$ are perfectly symmetric with respect to $\theta=\pi$) and in the high energy limit.  

This work represents a contribution towards a better understanding of the underlying mechanisms in the generation and distribution of the entanglement in the framework of fundamental interactions. A first extension of the present analysis, which is in progress, is represented by the detailed study of other basic QED scattering processes (M\"oller, Compton). We also plan to carry out the same investigation in different reference frames, also to test the Lorentz invariance of our results. Another important issue, which we have not considered in this work, is the study of tripartite entanglement in the output state: this will be also useful to understand the balance in the generation and transfer of the entanglement in the process, which does not appear to satisfy a simple sum rule.

\section{ACKNOWLEDGEMENTS}

M.B. wishes to thank Francesco Romeo for illuminating discussions. B.M. is greatful to Cristina Matrella, Gennaro Zanfardino, Pasquale Bosso and Gaetano Luciano for fruitful conversations on many topics related to the paper.

\section*{Appendix A}
\label{appA}

\textbf{Weil representation of $\gamma$-matrices}
\begin{align}
    \gamma^0 = 
    \begin{pmatrix}
        0 & \mathbb{1}\\
        \mathbb{1} & 0
    \end{pmatrix} , \hspace{0.3cm}
    \gamma^i = 
    \begin{pmatrix}
        0 & \sigma^i\\
        -\sigma^i & 0
    \end{pmatrix} , \hspace{0.3cm}
    \gamma^5 = 
    \begin{pmatrix}
        -\mathbb{1} & 0\\
        0 & \mathbb{1} 
    \end{pmatrix} .
\end{align}

Dirac spinors, as in Ref.\cite{Peskin:1995ev}, correspond to the particle and antiparticle solutions of the Dirac equations

\begin{equation}
    (\gamma^\mu p_\mu-m)u(p,s)=0,
\end{equation}
\begin{equation}
    (\gamma^\mu p_\mu+m)v(p,s)=0.
\end{equation}
They can be written as 
\begin{align}
     u(p,s)=\begin{pmatrix} \sqrt{p\cdot\sigma}\xi^s \\ \sqrt{p\cdot\bar\sigma} \xi^s 
     \end{pmatrix},\;
     v(p,s)=\begin{pmatrix} \sqrt{p\cdot\sigma} \xi^s \\ -\sqrt{p\cdot\bar\sigma} \xi^s \end{pmatrix},
\end{align} 
where the $\xi^s$ are the two component spinors eigenstates of helicity operator, $\sigma=(\mathbb{1},\vec{\sigma})$, $\sigma=(\mathbb{1},-\vec{\sigma})$ in which $\vec{\sigma}$ represents the Pauli matrices and $p=(\omega,\vec{p})$ is the four-momentum vector. The spinors below are expressed in terms of an arbitrary direction in which $\vec{p} = (\sin\theta\cos\phi,\sin\theta\sin\phi,\cos\theta)$, with $\theta, \phi$ the polar angles, and the subscrips $R$ and $L$ represent respectively the positive ($+1$) and negative $(-1)$ eigenvalues of the helicity operator.
\newpage

\textbf{Helicity Spinors}
\begin{align}
     {u_R(\vec{p})}=\begin{pmatrix} \sqrt{\omega-p} \cos{(\frac{\theta}{2})} \\ \sqrt{\omega-p}\hspace{0,2cm}e^{i\phi} \sin{(\frac{\theta}{2})} \\ \sqrt{\omega+p} \cos{(\frac{\theta}{2})} \\ \sqrt{\omega+p}\hspace{0,2cm}e^{i\phi} \sin{(\frac{\theta}{2})} \end{pmatrix},\;
     {u_L(\vec{p})}=\begin{pmatrix} -\sqrt{\omega+p} \sin{(\frac{\theta}{2})} \\ \sqrt{\omega+p}\hspace{0,2cm}e^{i\phi} \cos{(\frac{\theta}{2})} \\ -\sqrt{\omega-p} \sin{(\frac{\theta}{2})} \\ \sqrt{\omega-p}\hspace{0,2cm}e^{i\phi} \cos{(\frac{\theta}{2})} \end{pmatrix}
  \end{align}
     
 \begin{align}
     {v_R(\vec{p})}=\begin{pmatrix} -\sqrt{\omega+p} \sin{(\frac{\theta}{2})} \\ \sqrt{\omega+p}\hspace{0,2cm}e^{i\phi} \cos{(\frac{\theta}{2})} \\ \sqrt{\omega-p} \sin{(\frac{\theta}{2})} \\ -\sqrt{\omega-p}\hspace{0,2cm}e^{i\phi} \cos{(\frac{\theta}{2})} \end{pmatrix},\;
     {v_L(\vec{p})}=\begin{pmatrix} \sqrt{\omega-p} \cos{(\frac{\theta}{2})} \\ \sqrt{\omega-p}\hspace{0,2cm}e^{i\phi} \sin{(\frac{\theta}{2})} \\ -\sqrt{\omega+p} \cos{(\frac{\theta}{2})} \\ -\sqrt{\omega+p}\hspace{0,2cm}e^{i\phi} \sin{(\frac{\theta}{2})} \end{pmatrix}
  \end{align}

\begin{align}
     {u_R(-\vec{p})}=\begin{pmatrix} -\sqrt{\omega-p} \sin{(\frac{\theta}{2})} \\ \sqrt{\omega-p}\hspace{0,2cm}e^{i\phi} \cos{(\frac{\theta}{2})} \\ -\sqrt{\omega+p} \sin{(\frac{\theta}{2})} \\ \sqrt{\omega+p}\hspace{0,2cm}e^{i\phi} \cos{(\frac{\theta}{2})} \end{pmatrix},\;
     {u_L(-\vec{p})}=\begin{pmatrix} \sqrt{\omega+p} \cos{(\frac{\theta}{2})} \\ \sqrt{\omega+p}\hspace{0,2cm}e^{i\phi} \sin{(\frac{\theta}{2})} \\ \sqrt{\omega-p} \cos{(\frac{\theta}{2})} \\ \sqrt{\omega-p}\hspace{0,2cm}e^{i\phi} \sin{(\frac{\theta}{2})} \end{pmatrix}
  \end{align}
     
 \begin{align}
     {v_R(-\vec{p})}=\begin{pmatrix} \sqrt{\omega+p} \cos{(\frac{\theta}{2})} \\ \sqrt{\omega+p}\hspace{0,2cm}e^{i\phi} \sin{(\frac{\theta}{2})} \\ -\sqrt{\omega-p} \cos{(\frac{\theta}{2})} \\ -\sqrt{\omega-p}\hspace{0,2cm}e^{i\phi} \sin{(\frac{\theta}{2})} \end{pmatrix},\;
     {v_L(-\vec{p})}=\begin{pmatrix} -\sqrt{\omega-p} \sin{(\frac{\theta}{2})} \\ \sqrt{\omega-p}\hspace{0,2cm}e^{i\phi} \cos{(\frac{\theta}{2})} \\ \sqrt{\omega+p} \sin{(\frac{\theta}{2})} \\ -\sqrt{\omega+p}\hspace{0,2cm}e^{i\phi} \cos{(\frac{\theta}{2})} \end{pmatrix}
  \end{align}

\section*{Appendix B: Bhabha scattering amplitudes}
\label{appB}

The scattering amplitudes are calculated in the CM reference frame of particles $A$ and $B$. In the following, $p_1=(\omega,0,0,|\vec{p}|)$ and $p_2=(\omega,0,0,-|\vec{p}|)$ are the incoming 4-momenta that lie along the $z$-axis, while $p_3=(\omega,|\vec{p}|\sin\theta,0,|\vec{p}|\cos\theta)$ and $p_4=(\omega,-|\vec{p}|\sin\theta,0,-|\vec{p}|\cos\theta))$ are the outgoing four-momenta lying along a direction that form an angle $\theta$ with respect to $z$-axis. $a,b,r,s$ are the spin indices.
\begin{equation}
\mathcal{M}_{Bhabha}=e^2(\bar{v}(b,p_2)\gamma^{\mu}u(a,p_1)\frac{1}{(p_1+p_2)^2}\bar{u}(r,p_3)\gamma_{\mu}v(s,p_4)-\bar{v}(b,p_2)\gamma^{\mu}v(s,p_4)\frac{1}{(p_3-p_1)^2}\bar{u}(r,p_3)\gamma_{\mu}u(a,p_1))
\end{equation}

The explicit expressions for the polarized amplitudes are
\begin{equation}
\mathcal{M}(RR;RR)=\mathcal{M}(LL;LL)=\frac{(2+11\mu^2+8\mu^4+2\cos{\theta}+\mu^2\cos{2\theta})\csc^2{(\frac{\theta}{2})}}{4\mu^2(1+\mu^2)}
\end{equation}
%\vspace{0.5cm}
\begin{equation}
\mathcal{M}(RR;\prescript{RL}{LR}{})=-\mathcal{M}(LL;\prescript{RL}{LR}{})=-\frac{(1+\mu^2\cos\theta)\cot{(\frac{\theta}{2})}}{\mu^2\sqrt{1+\mu^2}}
\end{equation}
%\vspace{0.5cm}
\begin{equation}
\mathcal{M}(RR;LL)=\mathcal{M}(LL;RR)=\frac{1+\mu^2(1+\cos{\theta})}{\mu^2(1+\mu^2)}
\end{equation}
%\vspace{0.5cm}
\begin{equation}
\mathcal{M}(\prescript{RL}{LR}{};RR)=-\mathcal{M}(\prescript{RL}{LR}{};LL)=\frac{(1+\mu^2\cos\theta)\cot{(\frac{\theta}{2})}}{\mu^2\sqrt{1+\mu^2}}
\end{equation}
%\vspace{0.5cm}
\begin{equation}
\mathcal{M}(RL;RL)=\mathcal{M}(LR;LR)=\frac{(1+\mu^2(1+\cos{\theta}))\cot^2{(\frac{\theta}{2})}}{\mu^2}
\end{equation}
%\vspace{0.5cm}
\begin{equation}
\mathcal{M}(RL;LR)=\mathcal{M}(LR;RL)=1-\cos{\theta}-\frac{1}{\mu^2}
\end{equation}

\end{document}